\documentclass[prb,twocolumn]{revtex4}
\usepackage{graphicx}
\usepackage{color}
\usepackage{bm}
\usepackage{bbm}
\usepackage{dsfont}
\usepackage{mathbbol}
\usepackage{multirow}
\usepackage{float}
\usepackage{amsfonts}
\usepackage{amsmath}

\newcommand{\e}{\begin{equation}}
\newcommand{\ee}{\end{equation}}
\newcommand{\ea}{\begin{eqnarray}}
\newcommand{\eea}{\end{eqnarray}}

\newcommand{\ket}[1]{\left|#1\right\rangle}
\newcommand{\bra}[1]{\left\langle#1\right|}

\newcommand{\einheit}[1]{\hspace{1mm}\mbox{#1}}
\newsavebox{\mysquare}
\savebox{\mysquare}{\textcolor{black}{\rule{2.5mm}{2.5mm}}}

\begin{document}

\title{Electric $g$ Tensor Control and Spin Echo of a Hole-Spin Qubit in a
Quantum Dot Molecule}
\author{Robert Roloff}
\email{robert.roloff@uni-graz.at}
\affiliation{Fachbereich Theoretische Physik, Institut f\"{u}r Physik,
Karl-Franzens-Universit\"{a}t Graz, Universit\"{a}tsplatz 5, 8010 Graz,
Austria}
\author{Thomas Eissfeller}
\affiliation{Walter Schottky Institut, Technische Universit\"{a}t M\"{u}nchen, 85748
Garching, Germany}
\author{Peter Vogl}
\affiliation{Walter Schottky Institut, Technische Universit\"{a}t M\"{u}nchen, 85748
Garching, Germany}
\author{Walter P\"{o}tz}
\affiliation{Fachbereich Theoretische Physik, Institut f\"{u}r Physik,
Karl-Franzens-Universit\"{a}t Graz, Universit\"{a}tsplatz 5, 8010 Graz,
Austria}
\date{\today}

\begin{abstract}
The feasibility of high-fidelity single-qubit operations of a hole spin in a quantum dot molecule by electric $g$ tensor control is demonstrated. Apart from a constant external magnetic field the proposed scheme allows for an exclusively electric control of the hole spin. Realistic electric gate bias profiles are identified for various qubit operations using process-tomography-based optimal control. They are shown to be remarkably robust against decoherence and dissipation arising from the interaction of the hole with host-lattice nuclear spins and phonons, with a fidelity loss of $\approx$~1~percent for gate operation times of $\approx 10$~ns. Spin-echo experiments for the hole spin are modeled to explore dephasing mechanisms and the role of pulse-timing imperfections on the gate fidelity loss is discussed.\end{abstract}
\maketitle

\section{Introduction}
We propose and study feasibility of all-electric control of a qubit realization based on the hole spin in a quantum dot molecule. This system offers two main advantages over electron-spin-based realizations:  the use of hole spins increases the dephasing time associated with the interaction with nuclear spins by about an order of magnitude and allows for an efficient $g$ tensor control, thereby facilitating essentially all-electric control of the qubit.

Spin-based quantum bit realizations in semiconductor quantum dots have gained wide interest in the quantum computing community due to their potential regarding scalability and their long relaxation times. Most work, both theoretical and experimental, has focused on electron spin qubits, which have spin relaxation times ranging from several milliseconds~\cite{Kroutvar04} up to seconds.~\cite{Amasha08} However, it has turned out that this type of qubit is prone to decoherence processes due to the interaction with surrounding nuclear spins of the host lattice.~\cite{Coish05,Tay06} The Fermi contact hyperfine interaction ultimately limits the inhomogeneous dephasing time to $T_{2,e}^{\ast }\approx 10\hspace{1mm}\mbox{ns}$.~\cite{Petta05} Several techniques to circumvent this problem have been devised, e.g. nuclear state preparation,~\cite{Imamo03,Ramon07,Reilly08} fabrication of silicon-based semiconductor heterostructures with zero nuclear magnetic moment,~\cite{Eriksson04} or employing hole spins instead of electron spins.~\cite{Laurent05,Bulaev05} The $p$-type symmetry of the hole Bloch-function leads to a cancelation of the Fermi contact hyperfine interaction. However, it has been shown that the dipole-dipole hyperfine interaction and the coupling of the hole orbital angular momentum to the nuclear spins cannot be neglected.~\cite{Gryncharova77,Fischer08,Testelin09} These lead to an inhomogeneous dephasing time $T_{2,h}^{\ast }$ one order of magnitude longer than that of electrons, which is in good agreement with recent experiments.~\cite{Brunner09}

Basic spin qubit implementation schemes usually require local magnetic fields for individual spin manipulations.~\cite{DiVi98} Another technique, which has been proposed and tested recently, is to exploit the tunability of electron-spin and hole-spin $g$ tensors by electric fields rather than employing control by local magnetic fields.~\cite{Kato03,Doty06,Pingenot08,Andlauer09} If an external magnetic field $\vec{B}$ is present, the spin experiences an effective field $g(E)\cdot \vec{B}$ that can be changed locally by means of the electric field $E$. The dependence of the $g$ tensor on $E$ is particularly pronounced for hole spins in vertically stacked quantum dot molecules since the localization of the hole wave-function is very sensitive to externally applied electric fields.~\cite{Andlauer09} The goal of this paper is to show that such a system provides a fast and efficient universal gate for simple pulse shapes of $10\einheit{ns}$ duration.

In Sec.~\ref{sec:theory}, we describe the computational basis states of the qubit, the corresponding Hamiltonian and the mechanisms that lead to non-unitary dynamics. The equation of motion for the qubit, including decoherence and relaxation dynamics,  is given. In Sec.~\ref{sec:results}, we apply optimal control theory in order to find simple electric pulse shapes
that execute the Hadamard gate, as well as a $\pi /2$ and a $\pi $ pulse. We find that these transformations can be implemented with remarkably low fidelity losses of~$\approx 1\%$. In addition, we show how spin-echo experiments can be performed on the hole spin to explore decoherence in the system. We also discuss quantitatively the effect of pulse-timing imperfections. We conclude with a summary of the present work. The model for  the interaction of the nuclear spins with the hole is detailed  in Appendix~\ref{app:H1} and~\ref{app:H2}.

\section{Theory}
\label{sec:theory}
\begin{figure}[h]
\begin{tabular*}{8.6cm}{ll}
(a) & (b) \\
\includegraphics[width=4.3cm]{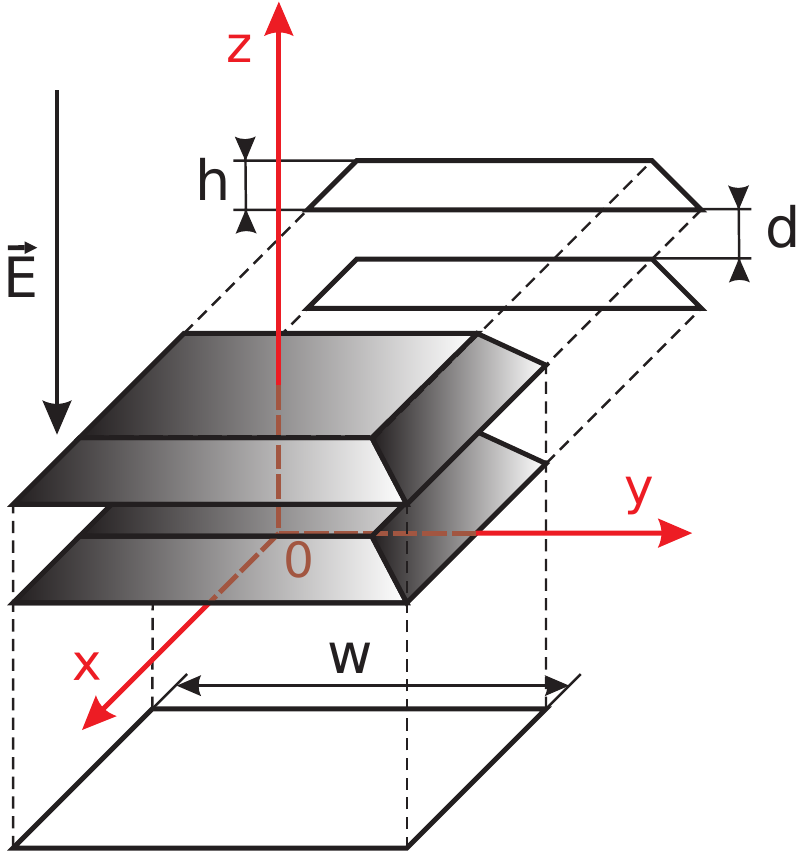} &
\raisebox{0.5cm}{\includegraphics[width=4cm]{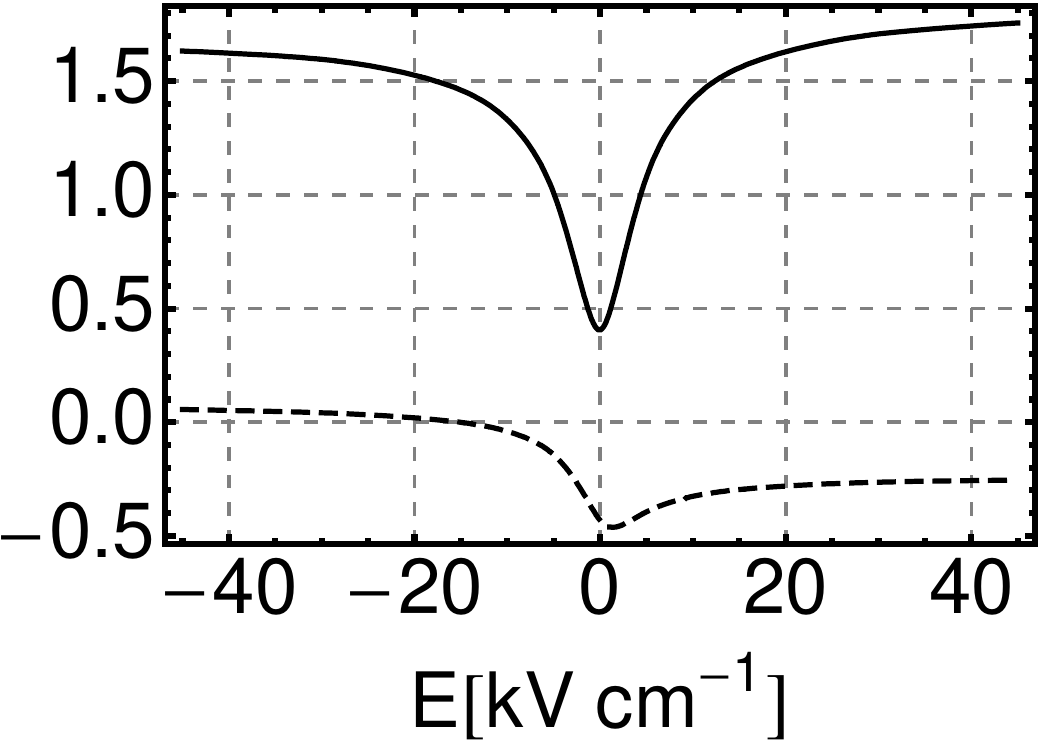}}
\end{tabular*}
\caption{(a) Vertically stacked pyramidal quantum dots, dot spacing $d=1.5\hspace{1mm} \mbox{nm}$, height $h=2.5\hspace{1mm}\mbox{nm}$, width $w=15\hspace{1mm}\mbox{nm}$. The coordinates $x$, $y$ and $z$ denote the $[100]$, $[010]$ and $[001]$ directions, respectively. The origin of the coordinate system is indicated by $0$. The external electric field points along the [001] direction, \textit{i.e.} $\vec{E}=(0,0,E)$. (b) Electric field dependence of the $g$ tensor elements. The solid and the dashed line denote $g^{001}(E)$ and $g^{1\bar{1}0}(E)$, respectively. (Data taken from Ref.~\onlinecite{Andlauer09}.)}
\label{g110andg001}
\end{figure}

In this section, we present the qubit realization and give the effective Hamiltonian for a single hole in a quantum dot molecule (QDM) composed of vertically-stacked self-assembled InAs/GaAs quantum dots (QD) separated by a small tunnel layer. The detailed geometry of the QDM, which we use for the present study, is shown in Fig.~\ref{g110andg001}(a). Two pyramidal dots (height $h$, width $w$) are stacked on top of each other and separated by a barrier of distance $d$. 
The QDM is exposed to a static magnetic field, as well as a time-dependent electric field applied along the [001] growth direction. It is controlled by a gate bias and used to modulate the hole $g$ factor by shaping of the hole wave function.~\cite{Andlauer09} Next to the interaction of the hole with the externally applied control fields, all contained in the effective Hamiltonian $H_0$, there will be additional (unwanted) interactions with the solid-state environment. 
Therefore, we account for the interaction of the host-lattice nuclear spins with the hole, develop a phenomenological description of the hole-phonon interaction, and give the resulting equation of motion on which our analysis is based.

\begin{figure}[h]
\begin{tabular*}{8.6cm}{lc}
(a) & hh envelope function $|F^{(1;3/2,-3/2)}(E,\vec r)|^2$ \\
& \\
& \includegraphics[width=7.5cm]{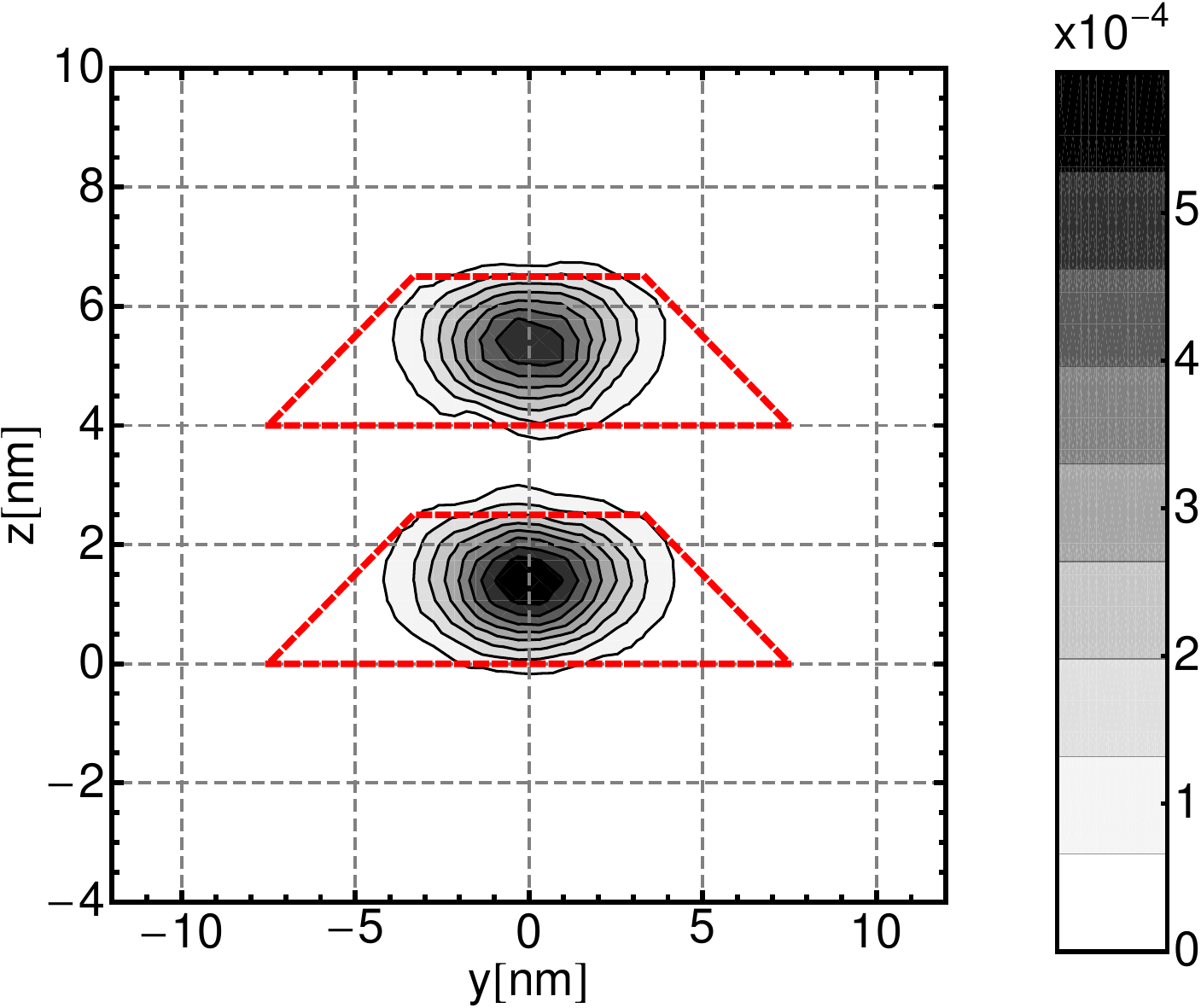} \\ 
&  \\ 
(b) & lh envelope function $|F^{(1;3/2,-1/2)}(E,\vec r)|^2$ \\
& \\
& \includegraphics[width=7.5cm]{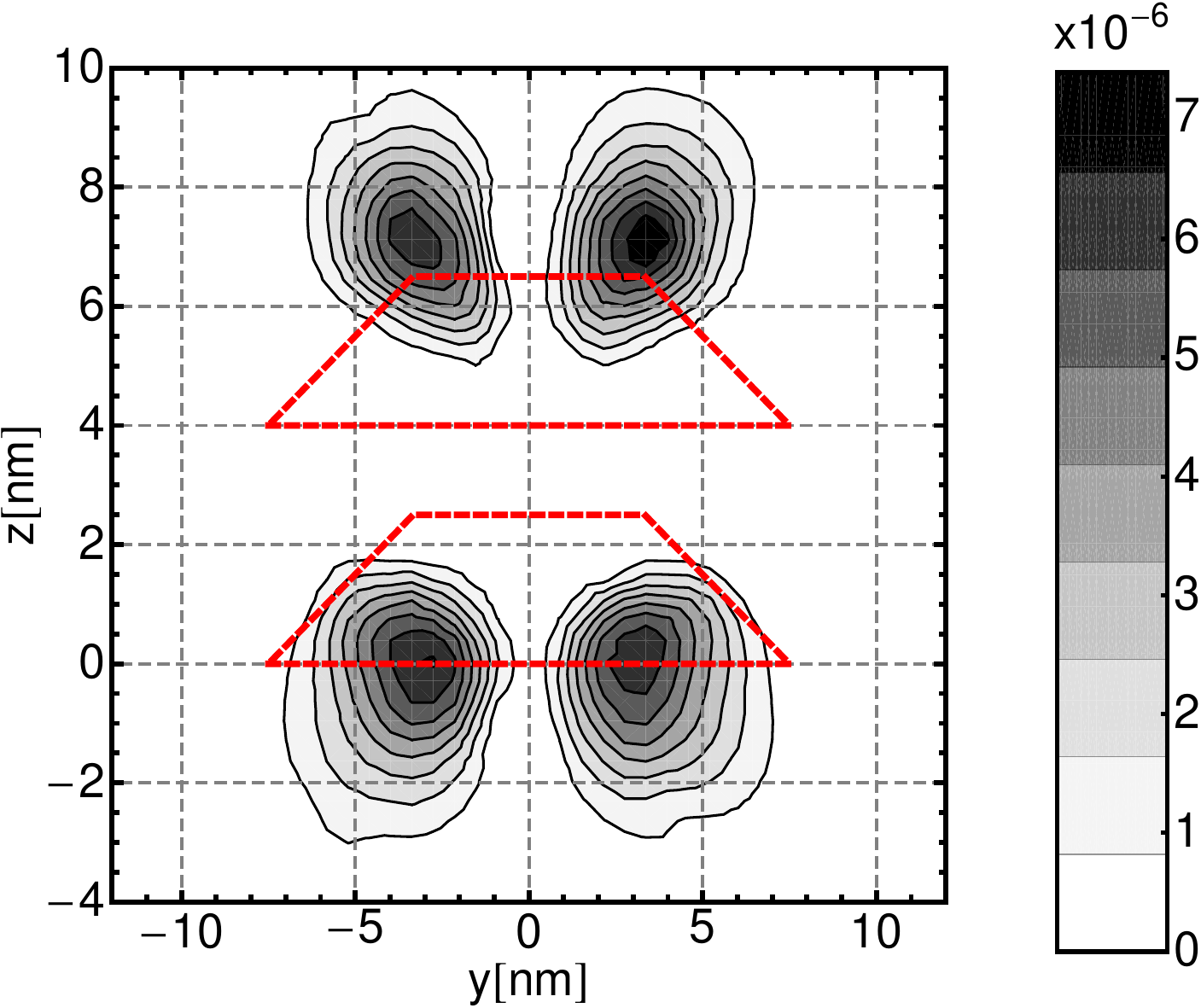}%

\end{tabular*}%
\caption{Spatial dependence of the envelope functions $|F^{(1;3/2;j_z)}(E, \vec r)|^2$ within the $x=0\einheit{nm}$ plane [see Fig.~\ref{g110andg001}(a)] for (a)~the hh contribution with $j_z=-3/2$ and (b)~for the lh contribution with $j_z=-1/2$. All envelope functions are given for $E=0\einheit{kVcm$^{-1}$}$ and $\vec B=(0,0,10)\einheit{mT}$. The cross sections of the pyramidal quantum dots are indicated by red dashed lines.}
\label{envelopes}
\end{figure}

The first task is to compute the effective interaction of the hole with a constant external magnetic field. It is characterized by a $g$ tensor, which depends on the externally applied electric field, as shown in Fig.~\ref{g110andg001}(b).~\cite{Andlauer09} We have performed three-dimensional 8-band envelope function calculations, including external fields, strain, and piezoelectric polarization, in order to determine the QDM heavy hole and light hole components of the ground $\left\vert \Psi _{0}\right\rangle $ and first excited Zeeman state $\left\vert \Psi _{1}\right\rangle $. Details of this method have been published elsewhere.~\cite{Andlauer08,Andlauer09b} Both states are predominantly heavy-hole (hh) like. However, light-hole (lh) contributions cannot be neglected. We write the hole wave functions as 
\begin{eqnarray}
&&\left\langle \vec{r}|\Psi _{k}\right\rangle \equiv \Psi _{k}(\vec{r}%
)=\sqrt{\Omega} \sum\limits_{j,j_{z}}{F^{(k;j,j_{z})}(E,\vec{r})\psi ^{(j,j_{z})}(\vec{r})}%
,  \notag \\
&&j\in \left\{ \frac{3}{2},\frac{1}{2}\right\} ,\;j_{z}\in \left\{ \pm \frac{%
3}{2},\pm \frac{1}{2}\right\} ,\;k\in \left\{ 0,1\right\} ,
\label{envelope_bloch}
\end{eqnarray}
where $F^{(k;j,j_{z})}(E,\vec{r})$ denotes the envelope function associated with the basis function $\psi ^{(j,j_{z})}(\vec{r})$, which transforms like the eigenfunction $\left\vert j,j_{z}\right\rangle $ of the angular momentum operator $J$, and $\Omega$ is the volume of the unit cell of the crystal. For zero electric field ${\vec E}$ and for vertical
magnetic field ${\vec B}=(0,0,10)\hspace{1mm}\mbox{mT}$, the hh and lh contributions, respectively, are given by 
\begin{eqnarray}
\int {d\vec{r}\;|F^{(k;3/2,\pm 3/2)}(E,\vec{r})|^{2}} &\approx &0.915,  \notag
\\
\int {d\vec{r}\;|F^{(k;3/2,\pm 1/2)}(E,\vec{r})|^{2}} &\approx &0.074.  \notag
\end{eqnarray}
We have chosen our coordinate system along the cubic axes, where $z$ is the growth direction of the QDs. The conduction and split-off (SO) band contributions are neglected, since they are smaller than $1\%$. In Fig.~\ref{envelopes}(a) and \ref{envelopes}(b), respectively,  we present contour plots of $|F^{(1;3/2,-3/2)}(E,\vec{r})|^{2}$ and $|F^{(1;3/2,-1/2)}(E,\vec{r})|^{2}$ in the $x=0$ plane. The corresponding coordinate system is depicted in Fig.~\ref{g110andg001}(a). 

The energy difference between $\left\vert \Psi _{1}\right\rangle $ and the next higher energy eigenstate $\left\vert \Psi _{2}\right\rangle $ is larger than $1\einheit{meV}$,  versus a splitting of ${\approx 0.4\einheit{$\mu$eV}}$ between the lowest two states. Hence, the system near the ground state is  well described by a two-level system with basis states $\left\vert \Psi _{0}\right\rangle $ and $\left\vert \Psi _{1}\right\rangle $. These are essentially linear combinations of upper- and lower-dot hole states, with the admixture depending upon the value of the external electric field. The effective Hamiltonian of the pseudo--spin system reads,~\cite{Andlauer09} 
\begin{equation}
H_{0}=\frac{\mu _{B}}{2}\vec{\sigma}\cdot g(E)\cdot 
\vec{B},  \label{H1}
\end{equation}%
where $\vec{\sigma}$, $g(E)$ and $\vec{B}$,  respectively,  denote the Pauli matrix vector of the pseudo spin-$\frac{1}{2}$ system, the electrically tunable hole $g$ tensor and the externally applied magnetic field.  Note that Eq.~\eqref{H1} is given in the basis $\left\{ \left\vert \Psi _{1}\right\rangle ,\left\vert \Psi _{0}\right\rangle \right\} $. By choosing a constant magnetic field of the form $\vec{B} =(B,-B,B)$, Eq.~\eqref{H1} takes the form
\begin{eqnarray}
H_{0} &=&\frac{\mu_B}{2}\vec{\sigma} 
\begin{bmatrix}
\frac{g^{1 1 0} + g^{1 \bar{1} 0}}{2} & \frac{g^{1 1 0} - g^{1 \bar{1} 0}}{2}
& 0 \\ 
\frac{g^{1 1 0} - g^{1 \bar{1} 0}}{2} & \frac{g^{1 1 0} + g^{1 \bar{1}0}}{2}
& 0 \\ 
0 & 0 & g^{001}%
\end{bmatrix}
\begin{bmatrix}
B \\ 
-B \\ 
B%
\end{bmatrix}
\notag \\
&=&\frac{\mu _{B}}{2}B\left[ (\sigma _{x}-\sigma _{y})g^{1\bar{1}%
0}+g^{001}\sigma _{z}\right].  \label{Hamiltonian1}
\end{eqnarray}
Throughout this paper, we choose a value of $B=10\hspace{1mm}\mbox{mT}$ and an external electric field pointing along the [001] direction, $\vec{E}=(0,0,E)$

\subsection{Hole Nuclear-Spin Interaction}
\label{sec:nuclear_interaction}
In addition to the externally applied magnetic field, the hole experiences an effective magnetic field that results from the nuclear spins of the host lattice. This is a consequence of the non-vanishing dipole-dipole hyperfine interaction (in contrast to the Fermi contact hyperfine interaction that vanishes for wave functions of $p$ symmetry), as well as the coupling of the hole orbital angular momentum to the nuclear spins.\cite{Fischer08,Testelin09} We now determine this effective field. 

The corresponding interaction Hamiltonian of a single nuclear spin with a hole has the form,~\cite{Abragam61} 
\begin{equation}
H_{\text{I}}^{i}=2\mu _{B}\gamma_i\vec{I}^{i}\cdot \left[ 
\frac{\vec{l}^{i}}{\rho _{i}^{3}}-\frac{\vec{s}}{\rho _{i}^{3}}+\frac{3\vec{\rho}%
_{i}\left( \vec{s}\cdot \vec{\rho}_{i}\right) }{\rho _{i}^{5}}\right] ,
\label{H_dipole-dipole}
\end{equation}%
where $\vec{I}^{i}$ denotes the $i$-th nuclear spin operator and $\gamma_i$ and $\mu_B$, respectively, denote the gyromagnetic ratio of the $i$-th nuclear spin and the Bohr magneton. The hole spin operator is denoted by $\vec{s}$, $\vec{\rho}_{i}$ is the distance vector between the hole and the $i$-th nuclear spin (located at $\rho=0$), and $\vec{l}^{i}=\vec{\rho}_{i}\times \vec{p}$ denotes the hole orbital angular momentum operator. The matrix elements of $H_{\text{I}}$ are now calculated in the $\left\{ \left\vert \Psi _{0}\right\rangle,\left\vert \Psi _{1}\right\rangle \right\} $ basis defined above. We closely follow the approach given in Refs.~\onlinecite{Gryncharova77} and~\onlinecite{Testelin09} and obtain 
\begin{equation}\label{matrix_elements}
\left\langle \Psi _{k}\right\vert H_{\text{I}}^{i}\left\vert \Psi
_{l}\right\rangle \equiv \sum\limits_{j_{z},j_{z}^{\prime }}{\left[ F^{(k;%
\frac{3}{2},j_{z}^{\prime })}(\vec{R}_{i})\right] ^{\ast }F^{(l;\frac{3}{2}%
,j_{z})}(\vec{R}_{i})V_{j_{z}^{\prime },j_{z}}^{i}},
\end{equation}%
where $\vec{R}_{i}$ denotes the position of the nuclear spin in the coordinate system of the envelope functions. The matrix elements with respect to the 4 basis functions $\left\{ \psi ^{(\frac{3}{2},\frac{3}{2})},\psi ^{(\frac{3}{2},\frac{1}{2})},\psi ^{(\frac{3}{2},-\frac{1}{2})},\psi^{(\frac{3}{2},-\frac{3}{2})}\right\} $ are given by~\footnote{We note that we find positive signs for the matrix elements $V_{1/2,-1/2}^{i}$ and $V_{-1/2,1/2}^{i}$ in contrast to Ref.~\onlinecite{Testelin09}.}
\begin{equation}
V_{j_{z}^{\prime },j_{z}}^{i}\equiv c_i
\begin{bmatrix}
I_{z}^{i} & \frac{1}{\sqrt{3}}I_{-}^{i} & 0 & 0 \\ 
\frac{1}{\sqrt{3}}I_{+}^{i} & \frac{1}{3}I_{z}^{i} & \frac{2}{3}I_{-}^{i} & 0
\\ 
0 & \frac{2}{3}I_{+}^{i} & -\frac{1}{3}I_{z}^{i} & \frac{1}{\sqrt{3}}%
I_{-}^{i} \\ 
0 & 0 & \frac{1}{\sqrt{3}}I_{+}^{i} & -I_{z}^{i}%
\end{bmatrix}
,
\end{equation}
where $I_{\pm }^{i}=I_{x}^{i}\pm I_{y}^{i}$ and $I_{z}^{i}$ are the nuclear spin operators and
\begin{equation}\label{radial}
c_i=\frac{8 \mu _{B} \gamma_i \hbar \Omega}{5}\int\limits_{0}^{R_{0}}{d\rho 
\frac{\left\vert \kappa(\rho)\right\vert ^{2}}{\rho}}.
\end{equation}
The integration in Eq.~\eqref{radial} extends over the dominant part of the interaction defined by a radius $R_{0}$ around the nuclear spin under consideration and $\kappa(\rho)$ is the radial part of the basis functions. The details of the calculation are given in Appendix~\ref{app:H1}. Finally, the interaction of the ensemble of nuclear spins with the two-level system can be cast into the form 
\begin{equation}
H_{\text{nuc}}(\vec{B}_{n})=\sum_{i}{H_{\text{I}}^{i}}=\frac{\mu _{B}}{2}%
\vec{\sigma}\cdot \vec{B}_{\text{n}},  \label{H_nuc_2}
\end{equation}%
where $i$ runs over all $N\approx 10^4 - 10^5$ nuclear spins interacting with the hole. This has the form of the interaction of the pseudo spin with an effective operator-valued magnetic field $\vec{B}_{\mathrm{n}}$. The dynamics of the hole is much faster than that of the nuclear spins. This allows one to employ a quasi-static approximation for $\vec{B}_{\text{n}}$ for the time period of a single measurement (initialization, manipulation and readout) of the hole spin qubit so that $\vec{B}_{\text{n}}$ may be approximated by a classical constant vector.~\cite{Merkulov02,Coish05,Tay06} During  the time it takes to perform  $10^{3}-10^{4}$ repetitions of the measurement, the effective nuclear magnetic field varies significantly, leading to inhomogeneous-broadening-type dynamics.~\cite{Merkulov02} The simplest way to take into account this variation is to treat the vector components of $\vec{B}_{\text{n}} = (B_{\text{n}}^{x}, B_{\text{n}}^{y}, B_{\text{n}}^{z})$ as random variables with Gaussian probability distributions,~\cite{Merkulov02,Coish05,Tay06} 
\begin{eqnarray}
&&P(\vec{B}_{\text{n}}) =P(B_{\text{n}}^{x})P(B_{\text{n}%
}^{y})P(B_{\text{n}}^{z}),  \label{prob_distr} \\
&&\mbox{with}\quad P(B_{\text{n}}^{i})=\frac{1}{\sqrt{2\pi }\Delta_{i}} \exp \left[ -(B_{\text{n}}^{i})^{2}/(2\Delta _{i}^{2})\right]. 
\notag
\end{eqnarray}
Here, $P(B_{\text{n}}^{i})$ is the probability of finding a value $B_{\text{n}}^{i}$ for the effective nuclear magnetic field along the $i$-direction and $\Delta _{i}\equiv \left\langle B_{\text{n}}^{i}B_{\text{n}}^{i}\right\rangle -\left\langle B_{\text{n}}^{i}\right\rangle ^{2}$ denotes the corresponding variance of the effective magnetic field fluctuation. We assume an ``infinite temperature'' nuclear-spin density matrix, \textit{i.e.} $\rho _{\text{n}}=\left( 2I+1\right) ^{-N}\openone$.~\cite{Tay06} Hence, the mean values $\left\langle B_{\text{n}}^{i}\right\rangle $ vanish for all directions $i$. Furthermore, we assume that the spin bath is uncorrelated,  \textit{i.e.} $\left\langle B_{\text{n}}^{i}B_{\text{n}}^{j}\right\rangle =0$. Finally, the total Hamiltonian of the system reads 
\begin{equation}
H(\vec{B}_{\text{n}})=H_{0}+H_{\text{nuc}}(\vec{B}_{\text{n}}).
\label{Hamiltonian2}
\end{equation}
We note that $H(\vec{B}_{\text{n}})$ is time-dependent via the electric field that is applied to control the hole dynamics.

In a recent experiment, the dephasing time of a hole spin in a single quantum dot has been measured as $T_{2,h}^{\ast }\approx 100\hspace{1mm}\mbox{ns}$.~\cite{Brunner09} This corresponds to a variance of $\Delta_{z}=\hbar/(\mu_B T_{2,h}^*) \approx 0.1\hspace{1mm}\mbox{mT}$.~\cite{Merkulov02} We use this value for our simulation of the hole nuclear-spin interaction. The variances $\Delta_{x}$ and $\Delta_{y}$ are calculated in Appendix~\ref{app:H2}. We find that $\Delta_{x}/\Delta_{z} \approx \Delta_{y}/\Delta_{z} \approx 10^{-1}$.

\subsection{Hole Phonon Interaction}
\label{sec:phonon_interaction}

The coupling of the hole to acoustic phonons via the piezoelectric and deformation potential interaction leads to additional dephasing and relaxation.~\cite{Lu05} To account for these mechanisms, we employ a Lindblad model.\cite{Lindblad76} The dissipator reads 
\begin{eqnarray}
\mathcal{D}\left[ \rho \right] =&&\Gamma_{\downarrow} \left[ \sigma _{-}\rho \sigma _{+}-\frac{1}{2}\left\{ \sigma _{+}\sigma _{-},\rho \right\} \right]+ \nonumber \\
&&\Gamma_{\uparrow} \left[ \sigma _{+}\rho \sigma _{-}-\frac{1}{2}\left\{ \sigma _{-}\sigma _{+},\rho \right\} \right]+\frac{\Gamma _{\text{ph}}^{\Phi }}{2}\left[ \sigma _{z}\rho \sigma _{z}-\rho \right],  \notag
\end{eqnarray}
with $\Gamma_{\downarrow}$, $\Gamma_{\uparrow}$ and $\Gamma _{\text{ph}}^{\Phi }$ denoting the relaxation rates for the transitions $\left\vert \Psi_{1}\right\rangle \rightarrow \left\vert \Psi _{0}\right\rangle $, $\left\vert \Psi_{0}\right\rangle \rightarrow \left\vert \Psi _{1}\right\rangle $ and the \textit{pure} dephasing rate, respectively. The braces denote anticommutators. For simplicity we set $\Gamma_{\text{ph}}^{\Phi }=\Gamma_\uparrow = \Gamma_\downarrow \equiv \Gamma $. The range of reported relaxation times varies significantly with  temperature and the externally applied magnetic field.~\cite{Heiss07,Gerardot08}
For this work, we choose a relatively conservative value of $T_{1,h}=1\hspace{1mm}\mbox{$\mu$s}=1/(\Gamma_\uparrow+\Gamma_\downarrow)=1/\Gamma $. We note that the hole relaxation time for quantum dots due to phonon interaction increases with decreasing external magnetic field.~\cite{Lu05,Bulaev05,Heiss07,Trif09}

\subsection{Hole Spin Dynamics}
\label{sec:hole_dynamics}

Inspection of Fig.~\ref{g110andg001}(b) reveals that the $g$ tensor component $g^{001}(E)$ cannot be tuned to~0, in contrast to $g^{1\bar{1}0}(E)$. For a simpler description of the hole-spin dynamics, it is therefore useful to switch to a rotating frame $\left\vert \psi \right\rangle \rightarrow \tilde{\left\vert \psi \right\rangle }\equiv U_{2}\left\vert \psi \right\rangle $
which rotates around the $z$-axis with a frequency given by 
\begin{equation}  \label{w*}
\omega ^{\ast}=\mu _{B}g^{001}(E^{\ast })B/\hbar.
\end{equation}
The transformation is characterized by the time dependent unitary operator $U_{2}\equiv \exp {\left[ i\omega ^{\ast }t\sigma _{z}/2\right] }$, where the electric field $E^{\ast }$ is defined by the relation 
\begin{equation}
g^{1\bar{1}0}(E^{\ast })=0.  \label{E*}
\end{equation}%
In addition, we perform another \textit{time independent} rotation $U_{1}=e^{-i\frac{\pi }{8}\sigma _{z}}$, which corresponds to the pseudo-spin rotation $\sigma _{x}-\sigma _{y}\rightarrow \sqrt{2}\sigma _{x}$ in Eq.~\eqref{Hamiltonian1}. In this rotating coordinate system (labeled by a tilde), the Lindblad equation for the density matrix reads
\begin{eqnarray}
\tilde{\rho}(\vec{B}_{\text{n}},t) &=&U_{2}U_{1}\rho (\vec{B}_{\text{n}%
},t)U_{1}^{\dag }U_{2}^{\dag },  \notag \\
\frac{d\tilde{\rho}(\vec{B}_{\text{n}},t)}{dt} &=&-\frac{i}{\hbar }\left[
H_{r}(\vec{B}_{\text{n}}),\tilde{\rho}(\vec{B}_{\text{n}},t)\right] +\tilde{%
\mathcal{D}}\left[ \tilde{\rho}(\vec{B}_{\text{n}},t)\right] ,  \notag \\
H_{r}(\vec{B}_{\text{n}}) &=&U_{2}U_{1}H(\vec{B}_{\text{n}})U_{1}^{\dag
}U_{2}^{\dag }+i\hbar \frac{dU_{2}}{dt}U_{2}^{\dag },  \label{dyn_open}
\end{eqnarray}
with $H(\vec{B}_{\text{n}})$ given in Eq.~\eqref{Hamiltonian2}. It can be shown that $\tilde{\mathcal{D}}\left[ \tilde{\rho}(\vec{B}_{\text{n}},t)\right] =\mathcal{D}\left[ \tilde{\rho}(\vec{B}_{\text{n}},t)\right] $, \textit{i.e.} the form of the dissipator is invariant under the coordinate transformations described above. The density matrix has to be averaged over the effective nuclear magnetic field $\vec{B}_{\text{n}}$ of each measurement.  It is calculated by averaging over typically $M=3000$ values of the random effective nuclear magnetic field with probability distributions as given in Eq.~\eqref{prob_distr}, using
\begin{equation}  \label{rho_1}
\tilde{\rho}(t)=\left( 1/M\right) \sum_{M}{\tilde{\rho}(\vec{B}_{\text{n}}).}
\end{equation}

\section{Results}
\label{sec:results}
\subsection{Pulse Shape Optimization}
\label{sec:optimization}
In the previous section we have detailed  the hole-spin qubit and identified a complete control mechanism.  We are now in a position to find optimal electric fields that perform any type of qubit transformation. However, the dependence of the $g$ tensor elements on the electric field $E$ is a complex one, as can be seen in Fig.~\ref{g110andg001}(b). Therefore, appropriate control fields can, in general, not be determined analytically, particularly, if the finite rise time of the electric control is to be taken into account. Here, we apply optimal control theory in order to determine both realistic and simple pulse shapes.

We start by outlining how to characterize qubit operations. For single qubit systems, quantum gate transformations can be described as rigid rotations of the Bloch sphere. For unitary dynamics, this rotation can be fully described by the time evolution propagator $U(t)$ corresponding to the Hamiltonian $H_{r}(\vec{B}_{\text{n}})$ given in Eq~\eqref{dyn_open}. However, it is more advantageous to employ the so-called \textit{process tomography matrix} (PTM) $\chi (t)$. For strictly unitary dynamics, this matrix takes the simple form,~\cite{Havel03} 
\begin{equation}
\chi (t)={U(t)}^{\ast }\otimes U(t),  \label{PTM_unitary}
\end{equation}
with ${U(t)}^{\ast }$ denoting the complex conjugate of the matrix representation of $U(t)$. The PTM eliminates physically irrelevant global phases of the
propagator $U\rightarrow e^{i\Phi }U$, can be readily obtained in experiments~\cite{OBrien04,Neeley08,Bialczak09}, and can be employed for non-unitary dynamics as well.~\cite{Poya97,Alt03}

The next step is to define a cost functional that quantitatively reflects how accurately a control field achieves a given unitary gate transformation. We use the PTM of Eq.~\eqref{PTM_unitary} that implicitly depends on the electric field, $\chi (t)=\chi \lbrack E(t);t]$, and seek electric pulse shapes that minimize the following cost functional,~\cite{Roloff09b}
\begin{equation}
J[E]=\operatorname{tr}\left\{ \left[ \chi (t_{f})-\chi _{D}\right] \left[ \chi
(t_{f})-\chi _{D}\right] ^{\dag }\right\} ,
\end{equation}
where $\chi _{D}$ denotes the ideal PTM and the interval $(0,t_{f})$ is the time-span allowed for the gate transformation (gate operation time). A perfect implementation of the given unitary gate transformation corresponds to $J=0$. Other distance measures for open quantum systems are given in Refs.~\onlinecite{Schulte06,Grace07,Wenin08c,Rebentrost09,Grace10}.
We choose the following analytic form of the electric control field,
\begin{eqnarray}
E(t) &=&E^{\ast }+\sum\limits_{j=1}^{M}{\ \frac{A_{j}}{4}\left( 1+\tanh {\left[ \alpha \left( t-\sum\limits_{i=1}^{j}\Delta {t_{i}}\right) \right] }\right) \times }  \notag \\
&&\left( 1+\tanh {\left[ \alpha \left( \sum\limits_{i=1}^{j}\Delta {t_{i}}-t\right) \right] }\right) ,  \label{E_opt}
\end{eqnarray}
where $A_{j}$ and $\Delta t_{j}$ are the parameters to be optimized and $E^{\ast }$ is the working point defined by Eq.~\eqref{E*}. The pulse form of Eq.~\eqref{E_opt} corresponds to a sequence of $M$ voltage steps of amplitude $A_{j}$, each of duration $\Delta t_{j}$. The finite rise time of the pulses is determined by the parameter $\alpha $ that we set to $\alpha
=8.79\einheit{ns$^{-1}$}$. The optimal control field $E_{\mathrm{opt}}(t)$ is obtained from the minimization procedure 
\begin{equation}\label{cost_func_minimization}	
J_{\text{opt}}[E_{\text{opt}}]=\min_{\left\lbrace A_{j},\Delta t_{j}\right\rbrace}J[E(A_{j},\Delta t_{j})].  
\end{equation}
Thus, a value $J > 0$ reflects suboptimal pulse shaping and/or the presence of decoherence and relaxation effects. In addition to $J$, it is customary to define a fidelity loss of the gate operation by 
\begin{equation}
\Delta F\left[ E\right] =\left( J\left[ E\right] /J_{\text{max}}\right) ^{\frac{1}{2}}
\end{equation}
with $J_{\text{max}}=2n^{2}$, where $n$ denotes the number of basis states of the quantum system, \textit{i.e.} $n=2$ for single qubits. A perfect execution of the gate operation corresponds to $\Delta F[E]=0$. For brevity, we will write $\Delta F$ instead of $\Delta F[E]$ throughout this work.

\begin{figure}[H]
\begin{tabular*}{8.6cm}{ll}
Hadamard gate &  \\ 
&  \\ 
(a) & \hspace{0.5cm} (b) \\ 
\includegraphics[width=5cm]{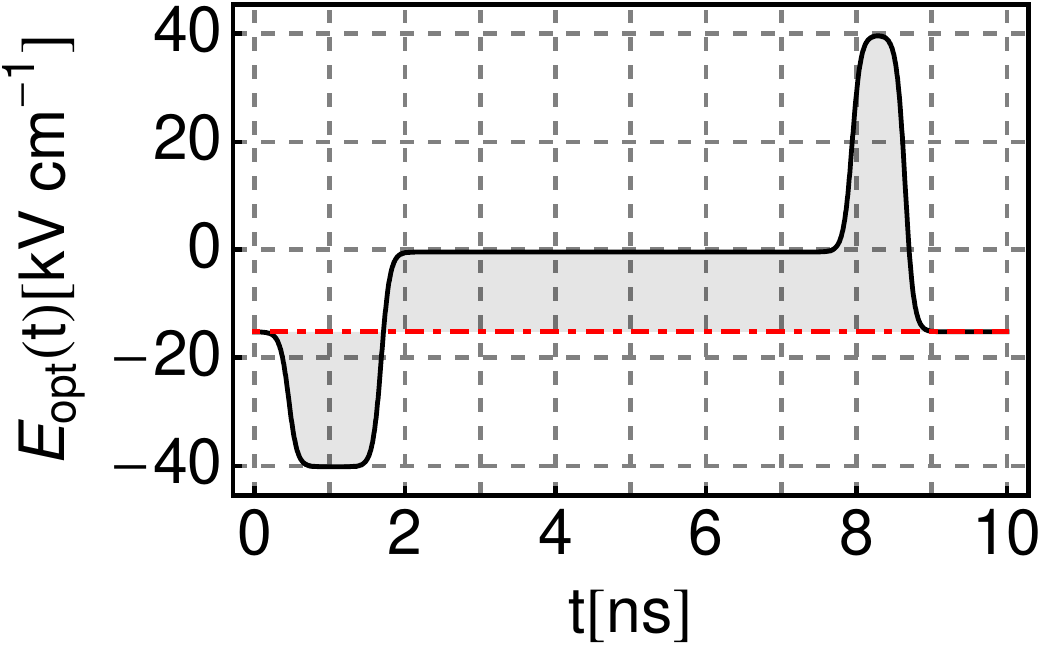} & \hspace{0.25cm}%
\includegraphics[width=3.2cm]{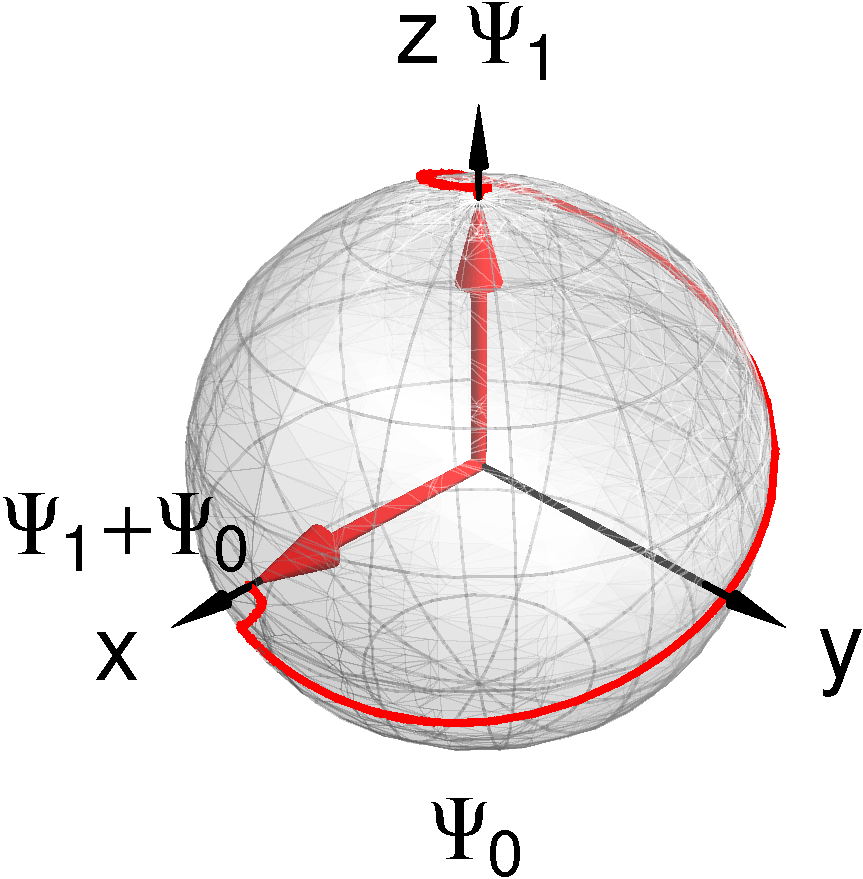} \\ 
&  \\ 
$\frac{\pi}{2}$ rotation &  \\ 
&  \\ 
(c) & \hspace{0.5cm} (d) \\ 
\includegraphics[width=5cm]{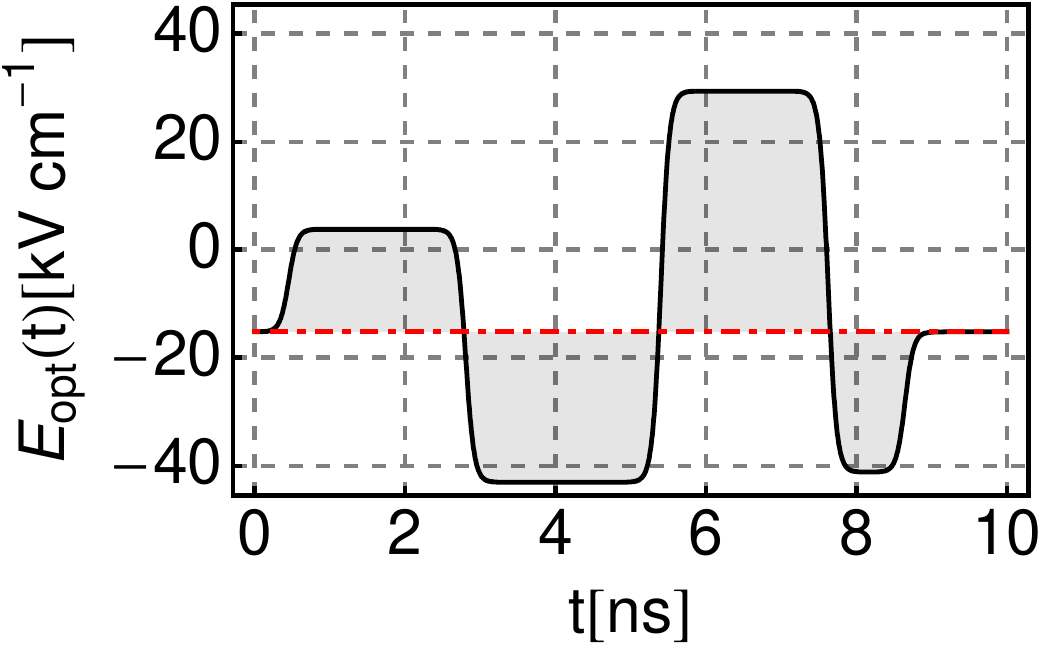} & \hspace{0.25cm}%
\includegraphics[width=3.2cm]{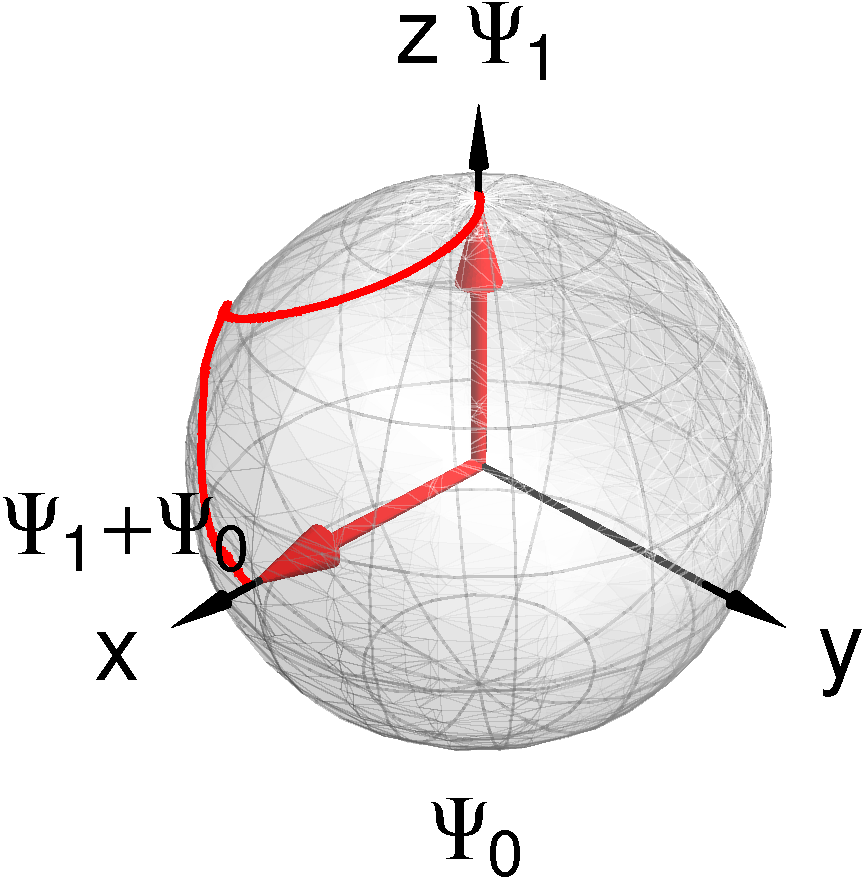}%
\end{tabular*}%
\caption{(a)~Optimal electric pulse shape $E_{\text{opt}}(t)$ for the Hadamard transformation and (c)~for the $\protect\pi /2$ rotation. The dot-dashed red lines correspond to $E=E^{\ast }$ given in Eq.~\eqref{E*}. (b)~The trajectory of the Bloch vector for the Hadamard gate and (d)~for the $\protect\pi /2$ pulse. The final states of both transformations coincide for
the initial state $\Psi_1$, though the trajectories of the Bloch vectors differ.}
\label{EPlot}
\end{figure}

\begin{figure}[h]
\begin{tabular*}{8.6cm}{ll}
(a) & \hspace{0.5cm} (b) \\ 
\includegraphics[width=3.9cm]{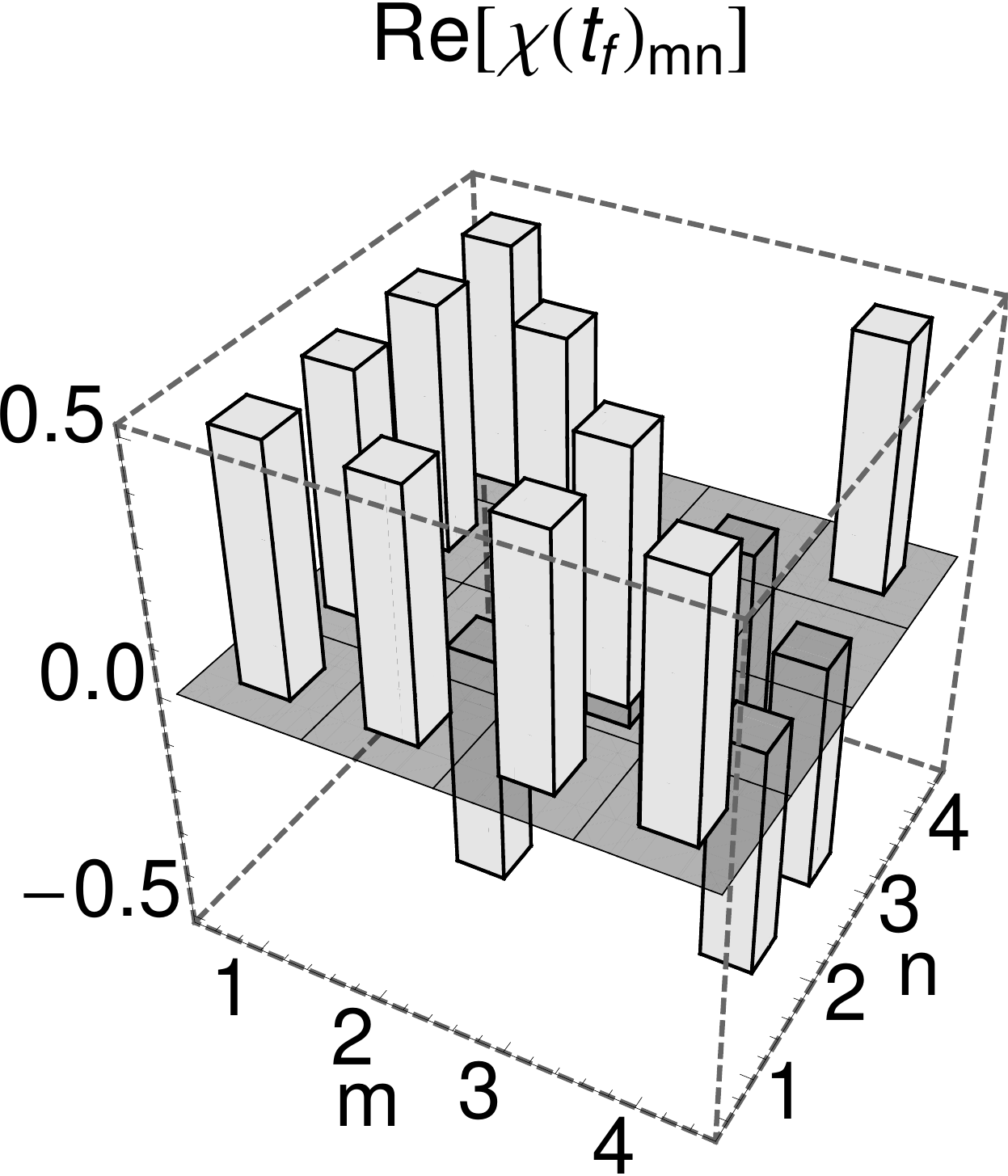} & \hspace{0.5cm} %
\includegraphics[width=3.8cm]{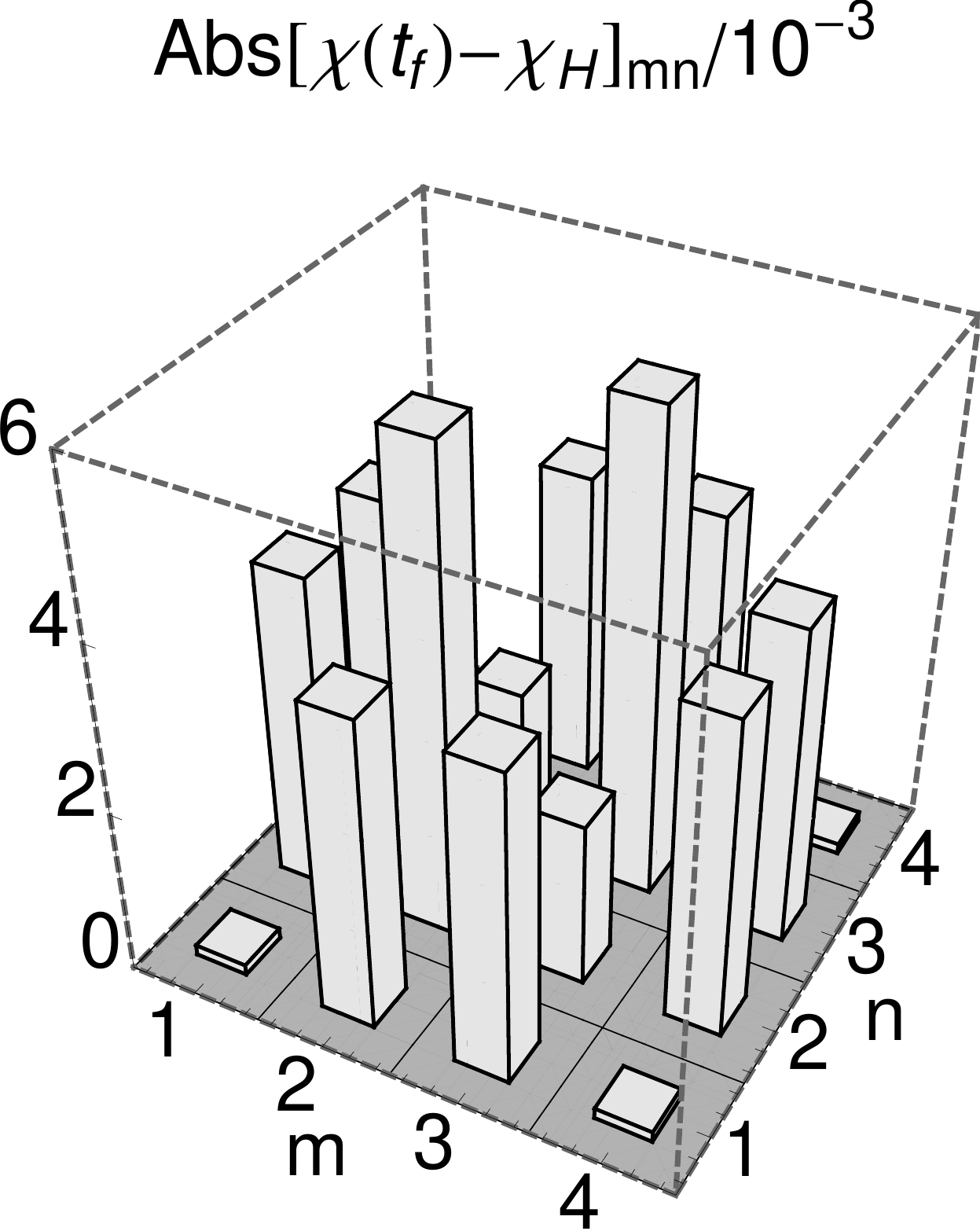}%
\end{tabular*}%
\caption{(a) Real part of the matrix elements $(m,n)$ of the optimal Hadamard PTM $\chi(t_f)$ (b) Deviation of the optimal PTM with respect to $\protect\chi_H$ (relaxation and dephasing included). The corresponding fidelity loss is $\Delta F \approx 1\%$.}
\label{ptms}
\end{figure}

We perform the cost functional minimization  Eq.~\eqref{cost_func_minimization} for both the Hadamard operation (which transforms the axes of the Bloch sphere as follows:  $x \rightarrow z$, $z \rightarrow x$ and $y\rightarrow -y$) and a $\pi /2$ rotation around the $y$-axis. The corresponding unitary evolution operators $U_{H}$ and $U_{\frac{\pi }{2}}$, and the ideal PTM $\chi _{H}$ and $\chi _{\frac{\pi }{2}}$, respectively, read [see Eq.~\eqref{PTM_unitary}]
\begin{eqnarray}
&&U_{H}=\frac{1}{\sqrt{2}}%
\begin{bmatrix}
1 & 1 \\ 
1 & -1%
\end{bmatrix}%
,\quad U_{\frac{\pi }{2}}=\frac{1}{\sqrt{2}}%
\begin{bmatrix}
1 & -1 \\ 
1 & 1%
\end{bmatrix}%
,  \notag \\
&&\chi _{H}=(U_{H})^{\ast }\otimes U_{H},\quad \chi _{\frac{\pi }{2}}=(U_{%
\frac{\pi }{2}})^{\ast }\otimes U_{\frac{\pi }{2}}.  \notag
\end{eqnarray}
We set the gate operation time $t_{f}=10\einheit{ns}$. The minimization is executed within a differential evolution algorithm.~\cite{Storn97} For the Hadamard gate, the algorithm converges to an optimal control field that consists of three pulses, depicted in Fig.~\ref{EPlot}(a). The $\pi /2$ rotation, on the other hand, is realized by an
optimal control field consisting of four pulses, as shown in Fig.~\ref{EPlot}(c). Trajectories of the Bloch vector for both transformations with equal initial states ($\Psi _{1}$) are given in Figs.~\ref{EPlot}(b) and~\ref{EPlot}(d).  If we neglect decoherence and relaxation, the fidelity losses due to imperfect pulse shaping are of the order $\Delta F\approx 0.001\%$.
When decoherence and relaxation are included, the fidelity loss increases to $\Delta F\approx 1\%$, which gives excellent performance. For the Hadamard gate, the real parts of the matrix elements $(m,n)$ of the optimal PTM, $\operatorname{Re}[\chi (t_{f})_{mn}]$, as well as the deviation of the optimal PTM from the ideal PTM $\chi _{H}$, are shown in Figs.~\ref{ptms}(a) and~\ref{ptms}(b), respectively.

\subsection{Hole Spin Echo}
\label{sec:spin_echo}
It is frequently important to analyze the characteristics of dephasing processes of a quantum system that are not associated with inhomogeneous dephasing, such as caused by the hole nuclear-spin interaction discussed in Sec. (A) in the present case. This can be achieved with spin-echo experiments.~\cite{Sli} The decay of the peak value of the spin echo recovery signal gives information about the additional cumulative dephasing rate and the time-dependence of the coherence loss (e.g. exponential versus polynomial decay).

\begin{figure}[h]
\begin{tabular*}{8.6cm}{ll}
(a) &  \\ 
&  \\ 
\multicolumn{2}{c}{\includegraphics[width=8cm]{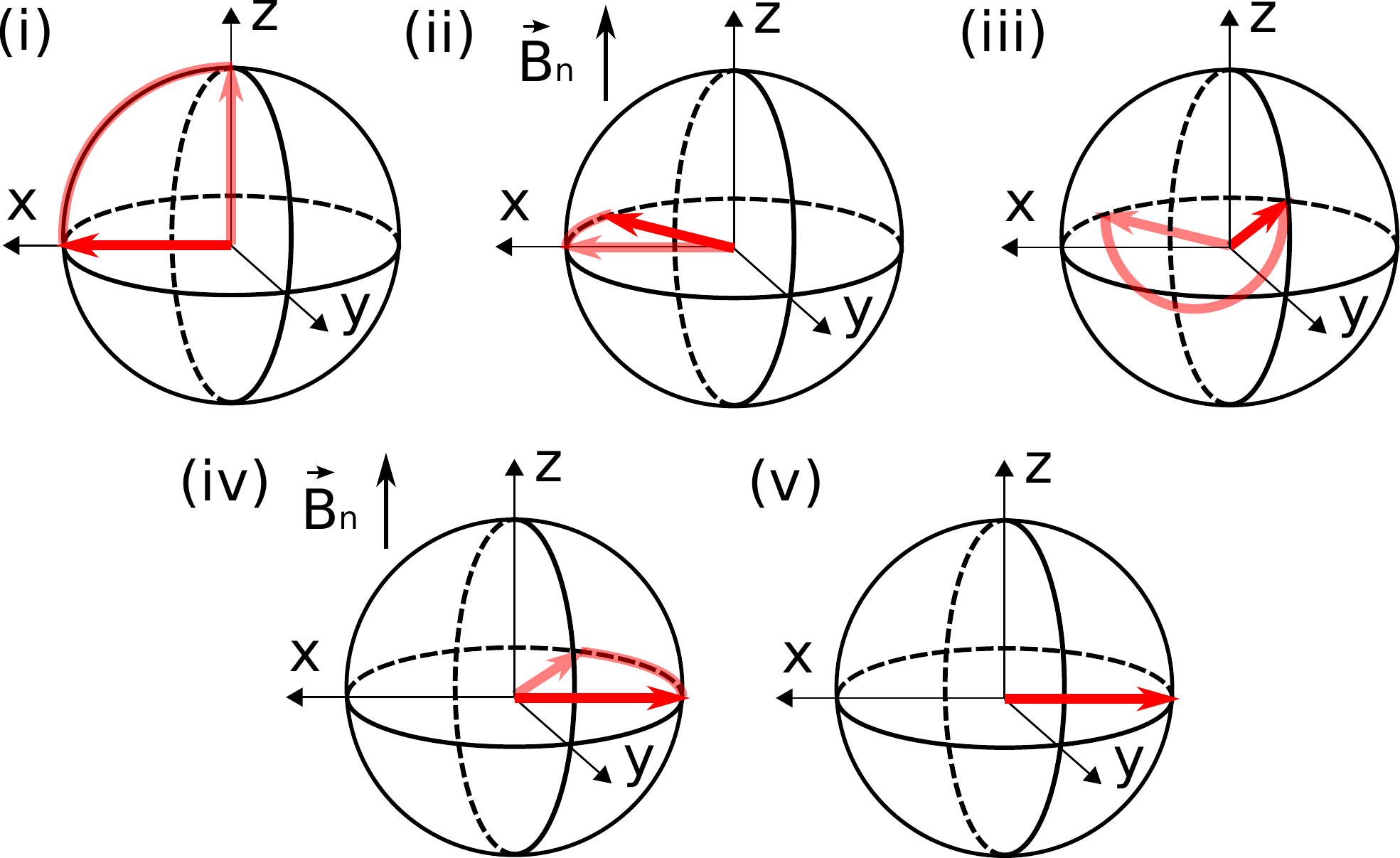}} \\ 
&  \\ 
(b) $\operatorname{Re}[\tilde \rho_{01}(t)]$ & (c) $\operatorname{Abs}[\tilde \rho_{01}(t)]$ \\ 
&  \\ 
\includegraphics[width=4.05cm]{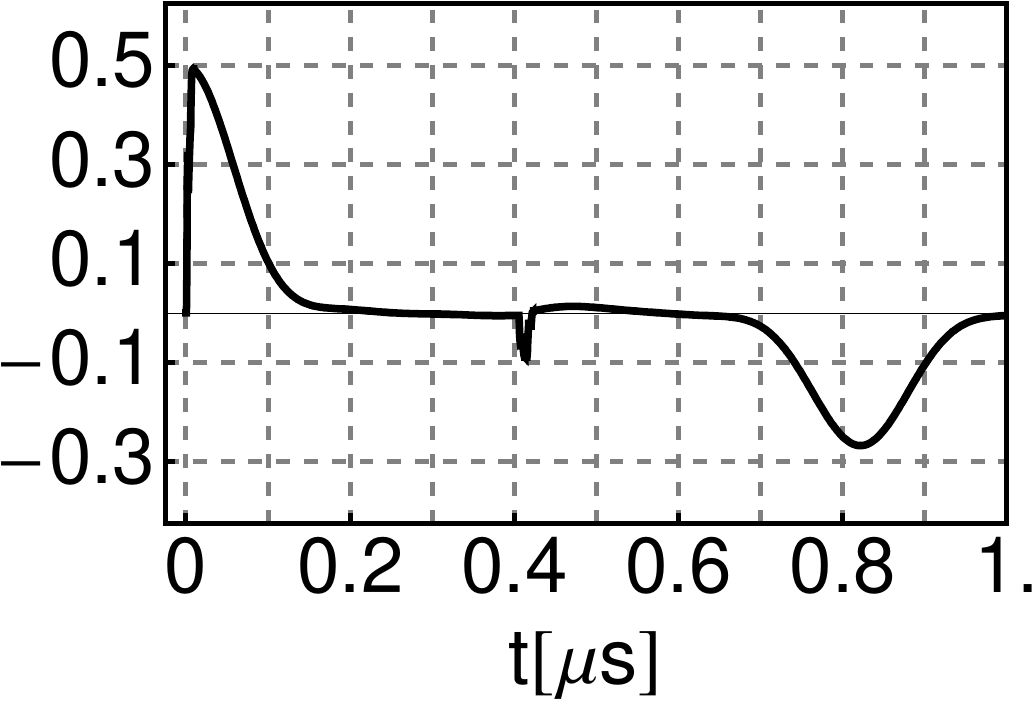} & \hspace{0.25cm} %
\includegraphics[width=3.95cm]{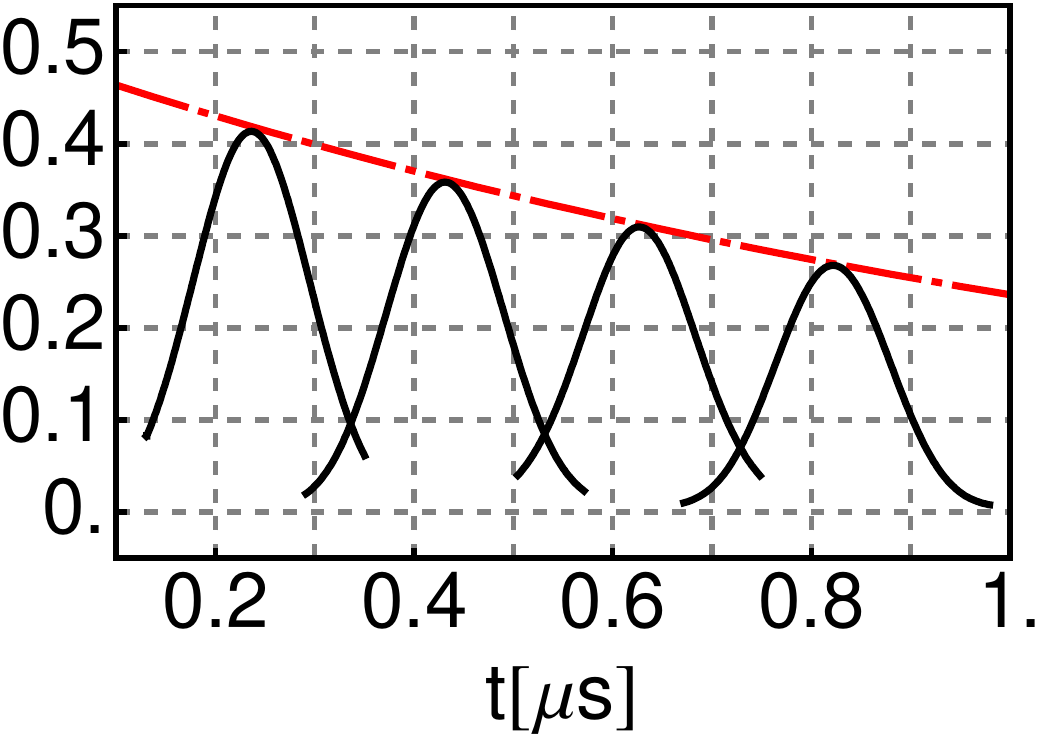}%
\end{tabular*}
\caption{(a)~Schematic spin-echo sequence as described in the main text. The red (dark gray) arrow denotes the Bloch vector. (b)~Time dependence of the coherence $\operatorname{Re}[\tilde \rho_{01}(t)]$. The spike at $t\approx 0.4\einheit{$\mu$s}$ is a consequence of the $\pi$-pulse. At $t\approx 0.8\einheit{$\mu$s}$ a partial revival of coherence can be observed. (c)~The solid black lines denote spin-echo signals for different pulse separation times $t_1$. The decrease of the peak echo signal is denoted by the red dot-dashed line.}
\label{SE}
\end{figure}

In Fig.~\ref{SE}(a), we show the crucial steps that are needed to perform a spin echo experiment for the hole-spin qubit.
\begin{enumerate}
\renewcommand{\labelenumi}{(\roman{enumi})} \setlength{\itemsep}{-2pt}
\item At time $t=0$, one applies a $\pi /2$ pulse such as given in Fig.~\ref{EPlot}(c). It transforms the initial qubit state $\left\vert \Psi_{1}\right\rangle $ into the superposition state $1/\sqrt{2}\left(\left\vert \Psi _{0}\right\rangle +\left\vert \Psi _{1}\right\rangle \right) $.~\cite{Sli}
\item Subsequently, the Bloch vector of the qubit evolves according to the system dynamics, including dissipation, decoherence and the effective magnetic field from the nuclear spins.
The latter causes a rotation of the Bloch vector around a random axis, given by the direction of $\vec{B}_{\text{n}}$.
\item At  time $t_{1}>0$ one applies a $\pi $ pulse that is composed of two subsequent $\pi /2$ pulses. It rotates the Bloch vector by an angle $\pi $ around the $y$-axis of the Bloch sphere.
\item Next, the Bloch vector evolves again and 
\item at time $t=2t_{1}$ the coherence, represented by $\operatorname{Re}[\tilde{\rho}_{01}(t)]$, is partly restored.
\end{enumerate}

Here, $\tilde{\rho}_{01}(t)$ is the off-diagonal matrix element of the density matrix $\tilde{\rho}(t)$ in the pseudo-spin basis [see Eq.~\eqref{rho_1}] and serves as a measure for coherence. For $t_{1}\approx 0.4\hspace{1mm}\mbox{$\mu$s}$, a plot of its time evolution during a spin-echo experiment, as obtained within our model, is shown in Fig.~\ref{SE}(b). Spin-echo signals corresponding to different pulse separation times $t_{1}$ are given in Fig.~\ref{SE}(c). Since relaxation and pure dephasing due to the spin-phonon interaction are included in our simulation, the restoration of coherence is obtained as imperfect. The peak values of the echoes decrease approximately as ${\sim \exp {\left\lbrace-2t_{1} \left[  \left( \Gamma_\uparrow + \Gamma_\downarrow\right)/2+\Gamma _{ph}^{\Phi }\right]\right\rbrace }  }$, indicated by the red dot-dashed line in Fig.~\ref{SE}(c). We note that the non-vanishing transversal nuclear magnetic field components $B_{n}^{x}$ and $B_{n}^{y}$ can also lead to a reduction of the peak value of the echo signal. However, we analyzed this effect and, for the present system,  found it to be negligible compared to both the longitudinal contributions $B_{n}^{z}$ and the phonon interaction.

\subsection{Pulse Timing Imperfections}
\label{sec:pulse_timing}
\begin{figure}[h]
\begin{tabular*}{8.6cm}{ll}
(a) & (b) \\ 
&  \\ 
\includegraphics[width=4.2cm]{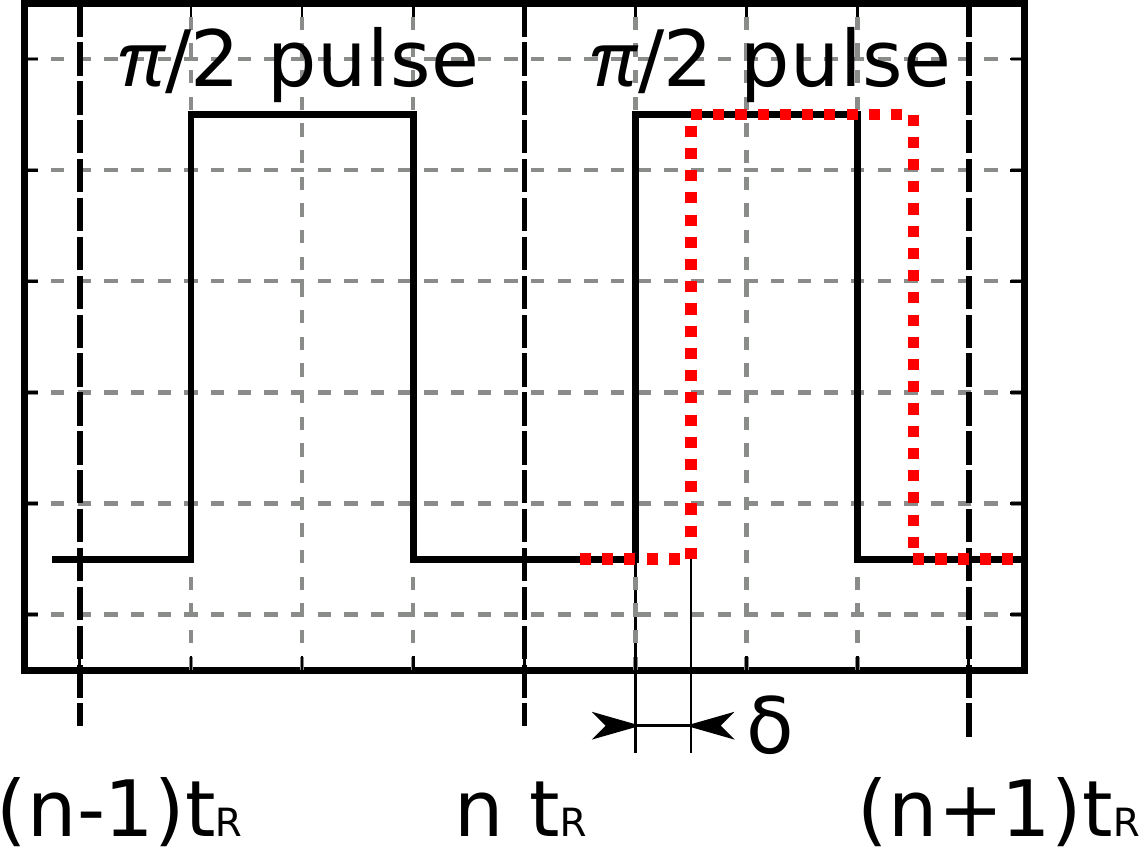} & \includegraphics[width=4.2cm]{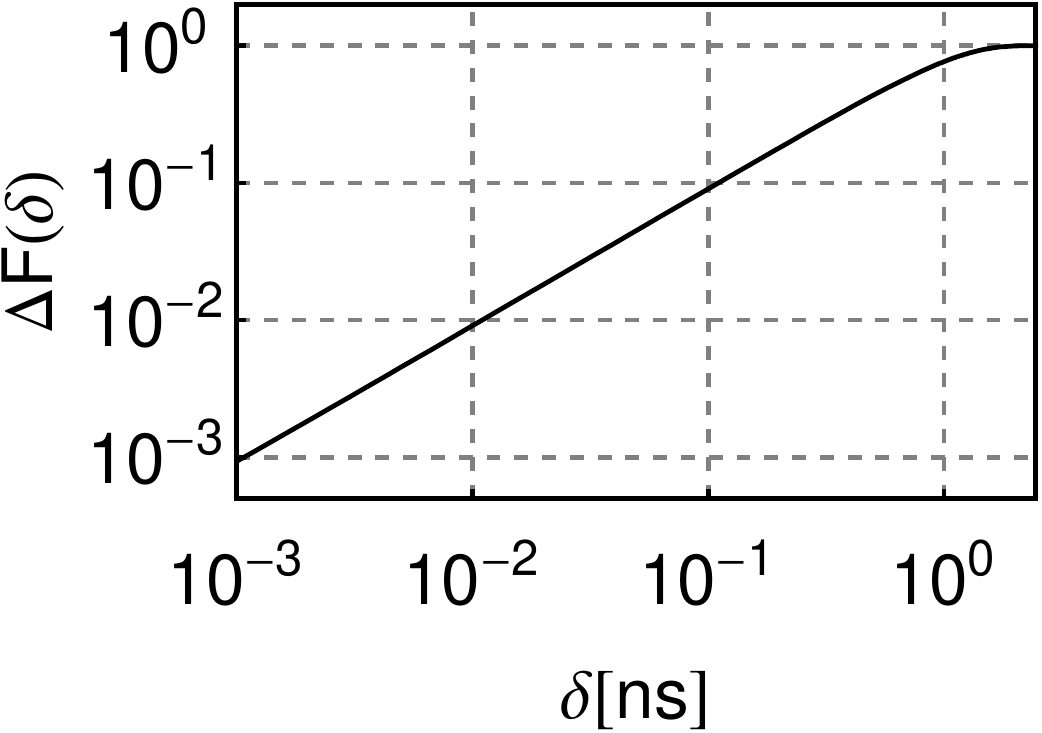}
\end{tabular*}
\caption{(a) The schematic ideal pulse sequence is denoted by the black solid line. The second imperfect $\protect\pi/2$ pulse (red/gray dashed line) is delayed by $\protect\delta$. (b) Double logarithmic plot of the fidelity loss $\Delta F$ vs. the delay time error $\protect\delta$.}
\label{robust}
\end{figure}

As mentioned above, the $\pi $-pulse of the proposed spin-echo sequence is composed of two subsequent $\pi /2$ pulses. Due to the form of the Hamiltonian in Eq.~\eqref{dyn_open}, these composite pulses should be applied at special times (integer multiples of $t_{R}=2\pi /\omega ^{\ast }$ after state initialization) in order to preserve the high fidelity of the gate transformation. 
This requirement is due to the effective magnetic field $g(E)\cdot \vec{B}$ along the $z$-direction that cannot be tuned to zero by electric means. The dependence of the fidelity loss $\Delta F$ with respect to a pulse delay time error $\delta $ between two subsequent $\pi /2$-pulses is shown in Fig.~\ref{robust}. This figure illustrates that the additional fidelity loss $\Delta F$ is smaller than $1\%$ as long as $\delta \leq 10\einheit{ps}$, which should be readily within reach of present day experiments. The sensitivity on the delay time error can be further decreased by reducing the magnitude of the external magnetic field.

\section{Summary}

\label{sec:summary} We propose a qubit realization in form of the spin of a hole in a quantum dot molecule, which is controlled all-electrically by $g$ tensor modulation, and evaluate its performance regarding controllability and dissipative effects. An effective qubit model is derived from a detailed electronic structure calculation, as well as the inclusion of the interaction with nuclear spins of the host lattice and phonons. On its basis, we predict that high-fidelity gate operations for a single hole within a quantum dot molecule are experimentally feasible and promising. The qubit can be fully controlled by means of electric $g$ tensor modulation. We use optimal control methods in order to determine simple electric control pulses, as illustrated for the Hadamard gate and for a $\pi/2$ qubit rotation around the $y$-axis of the Bloch sphere. The performance of these gate transformations is tested with respect to dephasing and relaxation due to the interaction of the hole spin with the surrounding nuclear spins of the host lattice and due to the hole-phonon interaction. For electric pulses of a duration of $10\einheit{ns}$, we find that  qubit manipulations can be performed with a remarkably low fidelity loss of $\Delta F\approx 1\%$. 
In addition, we propose a spin-echo experiment that allows one to completely cancel the inhomogeneous dephasing due to hole nuclear-spin interaction, as well as a study of additional dephasing mechanisms. 
We also investigate the influence of pulse-timing imperfections on the gate fidelity. We find that the additional fidelity loss is $\lesssim 1\%$ for pulse delay time errors of less than $10\einheit{ps}$.

\section*{Acknowledgment}

R.R. is grateful for helpful discussions with C. Testelin and U. Hohenester. The authors wish to acknowledge financial support of this work by FWF under Project No. P18829, as well as by the Deutsche Forschungsgemeinschaft Grants No. SFB 631 and No. SPP 1285 , and the Nanosystems Initiative Munich NIM.
\begin{widetext}
\begin{appendix}
\section{Matrix Elements of the Hole Nuclear-Spin Hamiltonian }
\label{app:H1}
In this section, we derive the matrix elements of Eq.~\eqref{matrix_elements}. For simplicity, we omit the nuclear spin index $i$. The interaction Hamiltonian of a nuclear spin $\vec I$ (located at the origin) with a hole (located at $\vec \rho$) reads,~\cite{Abragam61}
\begin{equation}\label{H_dd_1}
H_{\text{I}}(\vec \rho,\vec p)=2\mu_B \gamma \vec I \cdot \left[ \frac{ \vec l}{|\rho|^3}-\frac{\vec s}{|\rho|^3}+\frac{3 \vec \rho \left( \vec s \cdot \vec \rho\right) }{|\rho|^5}\right]=2\mu_B \gamma I_m \otimes \left[  \left(\frac{l_m}{|\rho|^3}\otimes \openone\right) - \left( \frac{\rho^2 \delta_{mn}-3\rho_m \rho_n}{|\rho|^5} \right)\otimes s_n \right],
\end{equation}
where we used the Einstein summation convention for the indices $m$ and $n$. Here $\gamma$, $\vec s$ and $\vec l=\vec \rho \times \vec p$, respectively, denote the gyromagnetic ratio of the nuclear spin $\vec I$, the hole spin and the hole orbital angular momentum. In Eq.~\eqref{H_dd_1}, we explicitly denoted the spatial and momentum dependence of $H_{\text{I}}$.
The hole wave function $\Psi$ can be written as a product of envelope functions $F$ and angular-momentum basis functions $\psi^{(j,j_z)}$,
\begin{equation}
\Psi_k(\vec r)=\sqrt{\Omega}\sum\limits_{j,j_z}{F^{(k;j,j_z)}(E,\vec r)\psi^{(j,j_z)}(\vec r)}, \quad j_z \in \left\lbrace \pm\frac{3}{2},\pm\frac{1}{2}\right\rbrace, \; j \in  \left\lbrace \frac{3}{2},\frac{1}{2}\right\rbrace, \; k \in  \left\lbrace 0,1\right\rbrace,
\end{equation}
or, using spin-resolved zone center valence-band Bloch functions $\psi^{(i,\alpha)}$, as
\begin{equation}
\Psi_k(\vec r)=\sqrt{\Omega}\sum\limits_{i,\alpha}{F^{(k;i,\alpha)}(E,\vec r)\psi^{(i,\alpha)}(\vec r)}, \quad i \in \left\lbrace X,Y,Z\right\rbrace, \; \alpha \in  \left\lbrace \uparrow,\downarrow\right\rbrace, \; k \in  \left\lbrace 0,1\right\rbrace.
\end{equation}
The matrix that describes the transformation between $\psi^{(j,j_z)}$ and $\psi^{(i,\alpha)}$ is given in Eq.~\eqref{basis_trafo}.~\cite{Andlauer09b,Ivchenko97}
The wave functions $\Psi_0(\vec r)$ and $\Psi_1(\vec r)$ denote the ground and the first excited Zeeman state of the quantum dot molecule. We neglect states with higher energy (see Sec.~\ref{sec:theory}) and calculate the matrix elements of the interaction Hamiltonian,
\begin{equation}
H_{\text{I}}(\vec \rho,\vec p) = \begin{bmatrix} 	\bra{\Psi_1} H_{\text{I}}(\vec \rho,\vec p) \ket{\Psi_1} & \bra{\Psi_1} H_{\text{I}}(\vec \rho,\vec p) \ket{\Psi_0} \\
								\bra{\Psi_0} H_{\text{I}}(\vec \rho,\vec p) \ket{\Psi_1} & \bra{\Psi_0} H_{\text{I}}(\vec \rho,\vec p) \ket{\Psi_0} \end{bmatrix},
\end{equation}
with								
\begin{equation}
\bra{\Psi_k} H_{\text{I}}(\vec \rho,\vec p) \ket{\Psi_l}=\Omega\sum\limits_{i,j,\alpha,\beta}{\int\limits_V{d \vec \rho d \vec \tau \left[ F^{(k;i,\alpha)}(E,\vec \tau+\vec R)\psi^{(i,\alpha)}(\vec \tau+\vec R)\right] ^*\bra{\vec \tau}H_{\text{I}}(\vec \rho,\vec p)\ket{\vec \rho} \psi^{(j,\beta)}(\vec \rho+\vec R) F^{(l;j,\beta)}(E,\vec \rho+\vec R)}},\nonumber
\end{equation}
where $\vec R$ denotes the position of the nuclear spin with respect to the hole envelope function coordinate system. Since the hole nuclear-spin interaction is short ranged, we can assume that it is non vanishing only within a sphere of volume $V_{R_0}$ and radius $R_0$ around the nuclear spin.~\cite{Gryncharova77} Furthermore, the envelope functions are approximately constant within this sphere, \textit{i.e.},
\begin{eqnarray}
\bra{\Psi_k} H_{\text{I}}(\vec \rho,\vec p) \ket{\Psi_l}&\approx& \Omega \sum\limits_{i,j,\alpha,\beta}{\left[ F^{(k;i,\alpha)}(E,\vec R)\right]^*F^{(l;j,\beta)}(E,\vec R)\int\limits_{V_{R_0}}{d \vec \rho \left[ \psi^{(i,\alpha)}(\vec \rho)\right] ^*H_{\text{I}}(\vec \rho,-i\hbar \vec \nabla_{\vec \rho})\psi^{(j,\beta)}(\vec \rho) }}, \nonumber \\
&\equiv & \sum\limits_{i,j,\alpha,\beta}{\left[ F^{(k;i,\alpha)}(E,\vec R)\right]^*F^{(l;j,\beta)}(E,\vec R)V_{ij\alpha\beta}},\label{H_dd_2} 
\end{eqnarray}
where we used the spatial periodicity of $\psi^{(i,\alpha)}(\vec \rho)$ and $\bra{\vec \tau}H_{\text{I}}(\vec \rho,\vec p)\ket{\vec \rho}\sim H_{\text{I}}(\vec \rho,-i\hbar \vec \nabla_{\vec \rho})\delta(\vec \rho - \vec \tau)$. In fact, it can be shown that $V_{R_0}$ is well approximated by the unit cell around the nuclear spin under consideration. Contributions due to long range interactions lead to corrections of the order of $1\%$.~\cite{Fischer08,Testelin09} Following Ref.~\onlinecite{Gryncharova77}, we employ a spherical approximation of the basis functions $\psi^{(i,\alpha)}(\vec \rho)$. Using spherical coordinates $(\rho,\theta,\phi)$ the corresponding approximate basis functions read,
\begin{eqnarray}
\psi^{(X,\alpha)}(\rho,\theta,\phi)&\approx& \sqrt{\frac{3}{4\pi}}\kappa(\rho)\sin\left( \theta\right) \cos\left( \phi\right)\ket{\alpha}, \nonumber \\
\psi^{(Y,\alpha)}(\rho,\theta,\phi)&\approx& \sqrt{\frac{3}{4\pi}}\kappa(\rho)\sin\left( \theta\right) \sin\left( \phi\right)\ket{\alpha}, \nonumber \\
\psi^{(Z,\alpha)}(\rho,\theta,\phi)&\approx& \sqrt{\frac{3}{4\pi}}\kappa(\rho)\cos\left( \theta\right)\ket{\alpha}. \nonumber
\end{eqnarray}
Here, $\kappa(\rho)$ denotes the radial part of $\psi^{(i,\alpha)}(\rho,\theta,\phi)$ and $\ket{\alpha}$ is a ket in the spin basis $\left\lbrace \uparrow, \downarrow\right\rbrace$. An estimation of the magnitude of the hole nuclear-spin interaction is given in Ref.~\onlinecite{Fischer08}, where the authors used a linear combination of atomic orbitals for the hh Bloch function.
In the basis $\psi^{(i,\alpha)}$ [$\alpha,\beta \in \left\lbrace \uparrow, \downarrow \right\rbrace$ and $i,j \in \left\lbrace X,Y,Z\right\rbrace $] and by employing the aforementioned spherical approximation, the integral $V_{ij\alpha\beta}$ [Eq.~\eqref{H_dd_2}] can be expressed as,
\begin{equation}\label{Vij}
V_{ij\alpha\beta}=2\mu_B \gamma\Omega\left[ \int\limits_{0}^{R_0}{d\rho\frac{\kappa(\rho)^2}{\rho}}\right] \left[ -i\hbar\epsilon_{ijm}I_m\delta_{\alpha\beta} - \frac{2}{5}\left( \delta_{ij}\delta_{mn}-\frac{3}{2}\left(\delta_{im}\delta_{jn}+\delta_{in}\delta_{jm} \right) \right) I_m \bra{\alpha}s_n\ket{\beta} \right].
\end{equation}
A discussion of modifications of the interaction at small $\rho $ due to the finite size of the nucleus can be found in Ref.~\onlinecite{Stoneham01}. 

Since we want to express the hole nuclear-spin interaction in terms of lh and hh contributions, we change the basis representation of $V_{ij\alpha\beta}$ from $\left\lbrace \psi^{(i,\alpha)}\right\rbrace $ to $\left\lbrace \psi^{(j,j_z)}\right\rbrace $. The transformation is governed by the unitary matrix,~\cite{Andlauer09b,Ivchenko97}
\begin{equation}\label{basis_trafo}
U_{\lambda\nu}=\frac{1}{\sqrt{6}}
\left[ \begin{array}{cccccc}
-\sqrt{3} & 0 & i\sqrt{3}& 0 & 0 & 0 \\
0 &-1 & 0 & i& 2 & 0 \\
1 & 0 & i& 0 & 0 & 2 \\
0 & \sqrt{3} & 0 & i\sqrt{3}& 0 & 0 \\
0 & -\sqrt{2} & 0 & i\sqrt{2} & -\sqrt{2} & 0 \\
-\sqrt{2} & 0 & -i\sqrt{2} & 0 & 0 & \sqrt{2}
\end{array}\right] ,,
\end{equation}
where the new matrix representation $\tilde V$ is given by $\tilde V_{\mu\nu}=U_{\mu\sigma} V_{\sigma \lambda} U^\dag_{\lambda\nu}$. Here, we use the shorthand notation
\begin{eqnarray}
&&\mu,\nu=(j,j_z) \in \left\lbrace \left(\frac{3}{2},+\frac{3}{2}\right), \left(\frac{3}{2},+\frac{1}{2}\right),\left(\frac{3}{2},-\frac{1}{2}\right), \left(\frac{3}{2},-\frac{3}{2}\right) \left(\frac{1}{2},+\frac{1}{2}\right), \left(\frac{1}{2},-\frac{1}{2}\right)\right\rbrace , \nonumber \\
&&\sigma,\lambda=(i,\alpha) \in \left\lbrace  (X,\uparrow),(X,\downarrow),(Y,\uparrow),(Y,\downarrow),(Z,\uparrow),(Z,\downarrow) \right\rbrace.
\end{eqnarray}
The transformed matrix $\tilde V_{\mu\nu}$ in the $\left\lbrace \psi^{(\frac{3}{2},+\frac{3}{2})}, \psi^{(\frac{3}{2},+\frac{1}{2})}, \psi^{(\frac{3}{2},-\frac{1}{2})}, \psi^{(\frac{3}{2},-\frac{3}{2})}\right\rbrace$ basis thus reads,
\begin{equation}
\tilde V_{\mu\nu}=\frac{8 \mu_B \gamma \hbar \Omega}{5}\int\limits_{0}^{R_0}{d\rho \frac{\left|\kappa(\rho)\right|^2}{\rho}}\begin{bmatrix} I_z & \frac{1}{\sqrt{3}}I_{-} & 0& 0 \\  \frac{1}{\sqrt{3}}I_{+} & \frac{1}{3}I_z& \frac{2}{3}I_{-} & 0 \\ 0 & \frac{2}{3}I_{+} & -\frac{1}{3}I_z & \frac{1}{\sqrt{3}}I_{-} \\ 0 & 0 & \frac{1}{\sqrt{3}}I_{+} & -I_z \end{bmatrix}\equiv \bar V_{j_z',j_z}. \label{V_def}
\end{equation}
Split-off contributions ($j=\frac{1}{2}$) have been neglected as discussed in Sec.~\ref{sec:theory}.
We finally arrive at the expression,
\begin{equation}
\bra{\Psi_k} H_{\text{I}}(\vec \rho,\vec p) \ket{\Psi_l}\approx \sum\limits_{j_z,j_z'}{\left[ F^{(k;j=3/2,j_z')}(E,\vec R)\right]^*F^{(l;j=3/2,j_z)}(E,\vec R)\bar V_{j_z',j_z}}, \label{H_fin}
\end{equation}
for the $\left\lbrace \ket{\Psi_0},\ket{\Psi_1}\right\rbrace$ matrix representation of the hole nuclear-spin interaction [for the definition of $\bar V_{j_z',j_z}$ see Eq.~\eqref{V_def}].

The Wigner-Eckart theorem can be used to determine the matrix elements of the hole nuclear-spin interaction as well.~\cite{Sak,Edmonds96,Cornwell84} In addition to the calculation presented above, we decomposed the spatial part of Eq.~\eqref{H_dd_1} into spherical tensors $\tilde l_m$ and $\tilde Q_{mn}$,
\begin{equation}
l_m/\rho^5 \rightarrow \tilde l_m, \quad \frac{\rho^2\delta_{mn}-3\rho_m\rho_n}{\rho^5} \rightarrow  \tilde Q_{mn}, \nonumber
\end{equation}
which correspond to angular momenta $j=1$ and $j=2$, respectively. The Wigner-Eckart theorem can then be readily applied to obtain the matrix elements in terms of Clebsch-Gordan coefficients and so-called \textit{reduced} matrix elements. However, in order to get a relation between the reduced matrix elements of the spherical tensors $\tilde l_m$ and $Q_{mn}$, approximations have to be employed as well. Hence, we decided to present the calculation of the matrix elements in terms of the spherical approximation given above.

\section{Variances of the Effective Nuclear Magnetic Field}
\label{app:H2}
Here, we derive the expression for the effective nuclear magnetic field $\vec B_n$ given in Eq.~\eqref{H_nuc_2}, as well as the ratio of the variances $\Delta_x$ and $\Delta_y$ with respect to $\Delta_z$ [Eqs.~\eqref{prob_distr}]. We rewrite Eq.~\eqref{H_fin} into,
\begin{equation}
\bra{\Psi_k} H_{\text{I}}(\vec \rho,\vec p) \ket{\Psi_l}=c \left[ A(E,\vec R) I_+ +A^\dag(E,\vec R) I_- + A^z(E,\vec R) I_z \right],
\end{equation}
with $A(E,\vec R)$, $A^\dag(E,\vec R)$ and $A^z(E,\vec R)$ denoting 2x2 matrices and with $c$ given in Eq.~\eqref{radial}. The corresponding matrix elements read,
\begin{eqnarray}
A_{kl}(E,\vec R)&=&\frac{1}{3}\left\lbrace  \sqrt{3} \left[ F^{(k;\frac{3}{2},-\frac{1}{2})}(\vec R)\right]^*  F^{(l;\frac{3}{2},-\frac{3}{2})}(\vec R)+2 \left[ F^{(k;\frac{3}{2},\frac{1}{2})}(\vec R)\right]^*  F^{(l;\frac{3}{2},-\frac{1}{2})}(\vec R) +\sqrt{3} \left[ F^{(k;\frac{3}{2},\frac{3}{2})}(\vec R)\right]^*  F^{(l;\frac{3}{2},\frac{1}{2})}(\vec R) \right\rbrace, \nonumber \\
A_{kl}^z(E,\vec R)&=&\frac{1}{3}\left\lbrace  -3 \left[ F^{(k;\frac{3}{2},-\frac{3}{2})}(\vec R)\right]^*  F^{(l;\frac{3}{2},-\frac{3}{2})}(\vec R) - \left[ F^{(k;\frac{3}{2},-\frac{1}{2})}(\vec R)\right]^*  F^{(l;\frac{3}{2},-\frac{1}{2})}(\vec R) +\right. \nonumber \\
&& \left.\left[ F^{(k;\frac{3}{2},\frac{1}{2})}(\vec R)\right]^*  F^{(l;\frac{3}{2},\frac{1}{2})}(\vec R) +3 \left[ F^{(k;\frac{3}{2},\frac{3}{2})}(\vec R)\right]^*  F^{(l;\frac{3}{2},\frac{3}{2})}(\vec R)\right\rbrace, \nonumber
\end{eqnarray}
where the electric field dependence of the envelope functions is omitted for brevity.

For the given dot size, the hole spin interacts with typically $N\approx10^4 - 10^5$ nuclear spins at positions $\vec R_i$ ($i=1...N$). The full interaction Hamiltonian thus reads (with nuclear spin index $i$),
\begin{equation}
H_{\text{nuc}} = \sum_{i}{H_{\text{I}}^i}=\sum_{i}{c_i\left\lbrace \left[ A(E,\vec R_i)+A^\dag(E,\vec R_i)\right] I_x^i + i\left[ A(E,\vec R_i)-A^\dag(E,\vec R_i)\right] I_y^i + A^z(E,\vec R) I_z^i \right\rbrace }. \label{Hamiltonian3}
\end{equation}
It can be cast into a pseudo spin-$\frac{1}{2}$ form, where the hole interacts with an effective nuclear magnetic field $\vec B_{\text{n}}$, \textit{i.e.}, $H_{\text{nuc}}= \frac{\mu_B}{2}  \vec \sigma \cdot \vec B_{\text{n}}$. Contributions proportional to the identity operator of the pseudo spin-$\frac{1}{2}$ space do not change the dynamics and have been omitted.
The $k$-th component of $\vec B_{\text{n}}$ is calculated by projecting Eq.~\eqref{Hamiltonian3} onto the Pauli matrix $\sigma_k$ ($k \in \left\lbrace x,y,z \right\rbrace$),
\begin{equation}
(\vec B_{\text{n}})_k = \frac{1}{\mu_B}\operatorname{tr}\left\lbrace \sigma_k H_{\text{nuc}}\right\rbrace \equiv \frac{1}{\mu_B}\sum_i{c_i\left[  g_{\text{n}}^{kx}(E,\vec R_i) I_x^i + g_{\text{n}}^{ky}(E,\vec R_i) I_y^i + g_{\text{n}}^{kz}(E,\vec R_i) I_z^i\right]  },
\end{equation}
with
\begin{equation}
g_{\text{n}}^{kx}(E,\vec R_i) \equiv  2 \operatorname{Re}\left[ \operatorname{tr}\left\lbrace \sigma_k A(E,\vec R_i)\right\rbrace \right], \;
g_{\text{n}}^{ky}(E,\vec R_i) \equiv  -2 \operatorname{Im}\left[ \operatorname{tr}\left\lbrace \sigma_k A(E,\vec R_i)\right\rbrace \right], \;
g_{\text{n}}^{kz}(E,\vec R_i) \equiv  \operatorname{tr}\left\lbrace \sigma_k A^z(E,\vec R_i)\right\rbrace. \nonumber
\end{equation}

As noted in Sec.~\ref{sec:nuclear_interaction}, we employ a quasi-static approximation for the dynamics of the nuclear spins. We further assume that the spins are uncorrelated, e.g. $\left\langle I_x^i I_x^{m} \right\rangle =0$ for $j\neq m$, and that only a single nuclear spin species with spin $I$ is present. 
The ``infinite-temperature'' density matrix of the nuclear spin ensemble reads $\rho_{\text{nuc}} = \openone (2I+1)^{-N}$. Hence, the mean values $\left\langle I_x^i\right\rangle$, $\left\langle I_y^i\right\rangle$ and $\left\langle I_z^i\right\rangle$ vanish and the variances of the nuclear spin components are $\left\langle I_x^i I_x^i\right\rangle=\left\langle I_y^i I_y^i\right\rangle=\left\langle I_z^i I_z^i\right\rangle=I(I+1)/3$. Furthermore, we can set $c_i=c$. The variances thus read
\begin{equation}\label{variances}
\Delta_k^2 = \left\langle B_{\text{n}}^k B_{\text{n}}^k \right\rangle = \left( \frac{c}{\mu_B}\right) ^2 \frac{I(I+1)}{3} \sum_i{\left\lbrace \left[ g_{\text{n}}^{kx}(E,\vec R_i)\right] ^2 + \left[ g_{\text{n}}^{ky}(E,\vec R_i)\right] ^2 + \left[ g_{\text{n}}^{kz}(E,\vec R_i)\right] ^2 \right\rbrace}.
\end{equation}
The sum in Eq.~\eqref{variances} is performed by a spatial sampling ($10^4$ points) of the lh and hh envelope functions. The value of the ratio of the variances with respect to $\Delta_z$ is $\Delta_x/\Delta_z \approx \Delta_y/\Delta_z \approx 0.12$. We then calculate $\Delta_x$ and $\Delta_y$ by setting $\Delta_z = 0.1\einheit{mT}$ (this value corresponds to the experimentally determined hole dephasing time $T_{2,h}^*\approx 100\einheit{ns}$, see Sec.~\ref{sec:nuclear_interaction}). We note that the effective nuclear magnetic field and the variances depend on the externally applied electric field via the electric-field dependence of the envelope functions. The values for $\Delta_k/\Delta_z$ given here have been calculated for $E=0\einheit{kVcm$^{-1}$}$ and $\vec B=(0,0,10)\einheit{mT}$.
\end{appendix}
\end{widetext}

\end{document}